\documentclass{aa}
\usepackage{epsfig,graphicx,natbib,url,twoopt,color}
\usepackage[varg]{txfonts}
\usepackage{hyperref}          %% for pdflatex
\usepackage{pdfcomment,acronym}      
\hypersetup{
  colorlinks=true,  
  urlcolor=blue,    
  linkcolor=red     
}
\def\# #1\par{\par\mbox{}\\ \noindent{\color{red}\small $\sharp$ #1}\\} 

\bibpunct{(}{)}{;}{a}{}{,}    %% natbib cite format used by A&A and ApJ
\makeatletter
 \newcommandtwoopt{\citeads}[3][][]{%
   \nonstopmode%              %% fix to not stop at error message in latex
   \href{http://adsabs.harvard.edu/abs/#3}%
        {\def\hyper@linkstart##1##2{}%
         \let\hyper@linkend\@empty\citealp[#1][#2]{#3}}%   %% Rutten, 2000
   \biblink{#3}{\href{http://adsabs.harvard.edu/abs/#3}{ADS}}%
   \errorstopmode}            %% fix to resume stopping at error messages 
 \newcommandtwoopt{\citepads}[3][][]{%
   \nonstopmode%              %% fix to not stop at error message in latex
   \href{http://adsabs.harvard.edu/abs/#3}%
        {\def\hyper@linkstart##1##2{}%
         \let\hyper@linkend\@empty\citep[#1][#2]{#3}}%     %% (Rutten 2000)
   \biblink{#3}{\href{http://adsabs.harvard.edu/abs/#3}{ADS}}%
   \errorstopmode}            %% fix to resume stopping at error messages
 \newcommandtwoopt{\citetads}[3][][]{%
   \nonstopmode%              %% fix to not stop at error message in latex
   \href{http://adsabs.harvard.edu/abs/#3}%
        {\def\hyper@linkstart##1##2{}%
         \let\hyper@linkend\@empty\citet[#1][#2]{#3}}%     %% Rutten (2000)
   \biblink{#3}{\href{http://adsabs.harvard.edu/abs/#3}{ADS}}%
   \errorstopmode}            %% fix to resume stopping at error messages 
 \newcommandtwoopt{\citeyearads}[3][][]{%
   \nonstopmode%              %% fix to not stop at error message in latex
   \href{http://adsabs.harvard.edu/abs/#3}%
        {\def\hyper@linkstart##1##2{}%
         \let\hyper@linkend\@empty\citeyear[#1][#2]{#3}}%  %% 2000
   \biblink{#3}{\href{http://adsabs.harvard.edu/abs/#3}{ADS}}%
   \errorstopmode}            %% fix to resume stopping at error messages 
\makeatother

\newcount\longrefs
\def\aap{\ifnum\longrefs=1 {Astron.\ Astrophys.}\else 
                           {A\hbox{\rm \&}A}\fi}
\def\aapr{\ifnum\longrefs=1 {Astron.\ Astrophys.\ Rev.}\else 
                            {A\hbox{\rm \&}AR}\fi}
\def\aaps{\ifnum\longrefs=1 {Astron.\ Astrophys.\ Suppl.}\else 
                            {A\hbox{\rm \&}A Suppl.}\fi}
\def\actaa{\ifnum\longrefs=1 {Acta Astronomica}\else
                            {Acta Astron.}\fi}
\def\aipcs{\ifnum\longrefs=1 {Am.\ Inst.\ Phys.\ Conf.\ Series}\else
                             {AIP Conf.\ Ser.}\fi}
\def\aj{\ifnum\longrefs=1 {Astron.\ J.}\else 
                          {AJ}\fi} 
\def\ao{\ifnum\longrefs=1 {Applied Optics}\else 
                           {Appl.\ Opt.}\fi} 
\def\aspcs{\ifnum\longrefs=1 {Astron.\ Soc.\ Pacific Conf.\ Series}\else 
                           {ASP Conf.\ Ser.}\fi} 
\def\apj{\ifnum\longrefs=1 {Astrophys.\ J.}\else 
                           {ApJ}\fi} 
\def\apjl{\ifnum\longrefs=1 {Astrophys.\ J. Lett.}\else 
                            {ApJL}\fi} 
\def\aplett{\ifnum\longrefs=1 {Astrophys.\ J. Lett.}\else 
                            {ApJ}\fi} 
\def\apjs{\ifnum\longrefs=1 {Astrophys.\ J. Suppl.}\else 
                            {ApJS}\fi}
\def\apss{\ifnum\longrefs=1 {Astrophys.\ and Space Science}\else 
                            {Astrophys.\ Space Sci.}\fi}
\def\araa{\ifnum\longrefs=1 {Ann.\ Rev.\ Astron.\ Astrophys.}\else 
                            {ARA\hbox{\rm \&}A}\fi}
\def\azh{\ifnum\longrefs=1 {Astronomicheskii Zhurnal}\else 
                            {Astron.\ Zhur.}\fi}
\def\baas{\ifnum\longrefs=1 {Bull.\ Am.\ Astron.\ Soc.}\else 
                            {BAAS}\fi}
\def\bain{\ifnum\longrefs=1 {Bull.\ Astronom.\ Institutes Netherlands}\else
                            {Bull.\ Astr.\ Inst.\ Neth.}\fi}
\def\cjaa{\ifnum\longrefs=1 {Chinese Jour.\ Astron.\ Astrophys.}\else 
                            {Chin.\ J.\ A\&A}\fi}
\def\gca{\ifnum\longrefs=1 {Geochim.\ Cosmochim.\ Acta}\else 
                           {Geochim.\ Cosmochim.\ Acta}\fi}
\def\grl{\ifnum\longrefs=1 {Geophys.\ Res.\ Lett.}\else 
                           {Geoph.\ Res.\ Lett.}\fi}
\def\iaucirc{\ifnum\longrefs=1 {IAU Circulars}\else 
                          {IAU Circ.}\fi}
\def\icarus{\ifnum\longrefs=1 {Icarus}\else 
                          {Icarus}\fi}
\def\ip{\ifnum\longrefs=1 {in press}\else 
                          {in press}\fi}
\def\jcap{\ifnum\longrefs=1 {Jour.\ Cosmology Astropart.\ Phys.}\else 
                          {JCAP}\fi}
\def\jgr{\ifnum\longrefs=1 {J.\ Geophys.\ Res.}\else 
                           {J.\ Geophys.\ Res.}\fi}  
\def\jrasc{\ifnum\longrefs=1 {J.\ Royal Astron.\ Soc.\ Canada}\else 
                           {JRAS Can.}\fi}  
\def\memsai{\ifnum\longrefs=1 {Mem.~Soc.~Astron.~Italiana}\else
                              {MmSAI}\fi}
\def\mnras{\ifnum\longrefs=1 {Mon.\ Not.\ Roy.\ Astron.\ Soc.}\else 
                             {MNRAS}\fi} 
\def\na{\ifnum\longrefs=1 {New Astronomy}\else 
                           {New Astron.}\fi}
\def\nar{\ifnum\longrefs=1 {New Astronomy rev.}\else 
                           {New Astron.\ Rev.}\fi}
\def\nat{\ifnum\longrefs=1 {Nature}\else 
                           {Nat}\fi}
\def\pasa{\ifnum\longrefs=1 {Pub.\ Astron.\ Soc.\ Australia}\else 
                            {PASA}\fi} 
\def\pasj{\ifnum\longrefs=1 {Pub.\ Astron.\ Soc.\ Japan}\else 
                            {PASJ}\fi} 
\def\pasp{\ifnum\longrefs=1 {Pub.\ Astron.\ Soc.\ Pacific}\else 
                            {PASP}\fi} 
\def\physscr{\ifnum\longrefs=1 {Physica Scripta}\else 
                            {Phys.\ Scrip.}\fi} 
\def\planss{\ifnum\longrefs=1 {Planetary \& Space Science}\else 
                            {Plan. \& Space Sci.}\fi} 
\def\procspie{\ifnum\longrefs=1 {Proc.\ SPIE}\else 
                            {Proc.\ SPIE}\fi} 
\def\qjras{\ifnum\longrefs=1 {Quarterly J.\ Royal Astron.\ Soc.}\else 
                            {QJRAS}\fi} 
\def\rmxaa{\ifnum\longrefs=1 {Revista Mexicana de Astron.\ y Astrofys.}\else 
                            {RMxAA}\fi} 
\def\sa{\ifnum\longrefs=1 {Soviet Astron..}\else 
                               {Sov.\ Astron.}\fi}
\def\skytel{\ifnum\longrefs=1 {Sky \& Telescope}\else 
                            {Sky \& Tel.}\fi} 
\def\solphys{\ifnum\longrefs=1 {Solar Phys.}\else 
                               {SoPh}\fi}
\def\sovast{\ifnum\longrefs=1 {Soviet Astronomy}\else 
                               {Sov.\ Ast.}\fi}
\def\ssr{\ifnum\longrefs=1 {Space Science Rev.}\else 
                               {Space\ Sci.\ Rev.}\fi}
\def\zap{\ifnum\longrefs=1 {Zeitschr.\ f.\ Astrophysik}\else
                               {Z.\ Astrophys.}\fi}

\makeatletter
\newcommand{\bibnote}[2]{\@namedef{#1note}{#2}}
\newcommand{\biblink}[2]{\@namedef{#1link}{#2}}
\makeatother

\def\acdef#1{\acl{#1} ({#1})}     %RR to avoid \ac first-use confusion
\newacro{AA}{Astronomy \& Astrophysics}
\newacro{ADS}{Astrophysics Data System}
\newacro{AIA}{Atmospheric Imaging Assembly}
\newacro{AO}{adaptive optics}
\newacro{ApJ}{Astrophysical Journal}
\newacro{AR}{active region}
\newacro{BFI}{Broad-band Filter Imager}
\newacro{CE}{coronal equilibrium}
\newacro{CfA}{Center for Astrophysics}
\newacro{CME}{coronal mass ejection}
\newacro{CRD}{complete redistribution}
\newacro{CRISP}{CRisp Imaging SpectroPolarimeter}
\newacro{CRISPEX}{CRisp SPectral EXplorer}
\newacro{CS}{coherent scattering}
\newacro{DEM}{Differential Emission Measure}
\newacro{DF}{dynamic fibril}
\newacro{DKIST}{Daniel K. Inouye Solar Telescope}
\newacro{DLR}{Deutsches Zentrum f\"ur Luft- und Raumfahrt}
\newacro{DOT}{Dutch Open Telescope}
\newacro{DST}{Richard B. Dunn Solar Telescope}   %RR or Hida domeless ST 
\newacro{EB}{Ellerman bomb}
\newacro{EDP}{\'{E}dition Diffusion Presse Sciences}  %RR they say so
\newacro{EIT}{Extreme ultraviolet Imaging Telescope}
\newacro{EPIC}{European participation in Solar-C}
\newacro{ERC}{European Research Council}
\newacro{ESA}{European Space Agency}
\newacro{EST}{European Solar Telescope}
\newacro{EUV}{extreme ultraviolet}
\newacro{FAF}{flaring active-region fibril}
\newacro{FITS}{Flexible Image Transport System}
\newacro{FOV}{field of view}
\newacro{fov}{field of view}
\newacro{FWHM}{full width at half maximum}
\newacro{HAO}{High Altitude Observatory}
\newacro{HD}{hydrodynamics}
\newacro{Hi-C}{High Resolution Coronal Imager Sounding Rocket}
\newacro{HMI}{Helioseismic and Magnetic Imager}
\newacro{IAA}{Instituto de Astrof\'{i}sica de Andaluc\'{i}a}
\newacro{IAC}{Instituto de Astrof\'{i}sica de Canarias}
\newacro{IAS}{Institut d'Astrophysique Spatiale}
\newacro{IDL}{Interactive Data Language}
\newacro{IMaX}{Imaging Magnetograph eXperiment}
\newacro{INAF}{Istituto Nazionale di Astrofisica}
\newacro{IB}{IRIS bomb}
\newacro{IR}{infrared}
\newacro{IRIS}{Interface Region Imaging Spectrograph}
\newacro{ISAS}{Institute of Space and Astronautical Science}
\newacro{ISP}{Institute for Solar Physics}
\newacro{ISS}{International Space Station}
\newacro{ISSI}{International Space Science Institute}
\newacro{ITA}{Institute for Theoretical Astrophysics}
\newacro{JAXA}{Japan Aerospace Exploration Agency}
\newacro{KIS}{Kiepenheuer--Institut f\"{u}r Sonnenphysik}
\newacro{KPNO}{Kitt Peak National Observatory}
\newacro{LASP}{Laboratory for Atmospheric and Space Physics}
\newacro{LC}{liquid cristal}
\newacro{LMSAL}{Lockheed Martin Solar and Astrophysics Labratory}
\newacro{LOS}{line of sight}
\newacro{LTE}{local thermodynamic equilibrium}
\newacro{MC}{magnetic concentration}
\newacro{MCAO}{multi-conjugate adaptive optics} 
\newacro{MDI}{Michelson Doppler Imager}
\newacro{ME}{Milne-Eddington} 
\newacro{MHD}{magnetohydrodynamics}
\newacro{MOMFBD}{Multi-Object Multi-Frame Blind Deconvolution}
\newacro{MPE}{Max--Planck--Institut f\"ur extraterrestrische Physik}
\newacro{MPG}{Max--Planck--Gesellschaft}
\newacro{MPS}{Max Planck Institute for Solar System Research}
\newacro{MSSL}{Mullard Space Science Laboratory}
\newacro{MTF}{modulation transfer function}
\newacro{NAOJ}{National Astronomical Observatory of Japan}
\newacro{NASA}{National Aeronautics and Space Administration}
\newacro{NLTE}{non-local thermodynamic equilibrium}
\newacro{NLFFF}{non-linear force-free field}
\newacro{NOAA}{National Oceanic and Atmospheric Administration}
\newacro{non-E}{non-equilibrium}
\newacro{NSO}{National Solar Observatory}
\newacro{NWO}{Netherlands Organisation for Scientific Research}
\newacro{PRD}{partial redistribution}
\newacro{PROBA2}{PRoject for OnBoard Autonomy}
\newacro{PSF}{point spread function}
\newacro{QS}{quiet Sun}
\newacro{RAL}{Rutherford Appleton Laboratory}
\newacro{RBE}{rapid blue-shifted excursion}
\newacro{R-MHD}{radiation hydrodynamics}
\newacro{rms}{root mean square}
\newacro{RMS}{root mean square}
\newacro{ROB}{Royal Observatory of Belgium}
\newacro{ROI}{region of interest}
\newacro{RRE}{rapid red-shifted excursion}
\newacro{RTE}{radiative transfer equation}
\newacro{SE}{statistical equilibrium}
\newacro{SB}{Saha Boltzmann}
\newacro{SDO}{Solar Dynamics Observatory}
\newacro{SJI}{slit-jaw image}
\newacro{SNR}{signal-to-noise ratio}
\newacro{SO}{Solar Orbiter}
\newacro{SoHO}{Solar and Heliospheric Observatory}
\newacro{SP}{Spectropolarimeter}
\newacro{SST}{Swedish 1-m Solar Telescope}
\newacro{SUMER}{Solar Ultraviolet Measurements of Emitted Radiation}
\newacro{SUFI}{Sunrise Filter Imager}
\newacro{SVD}{singular value decomposition}
\newacro{SVST}{Swedish Vacuum Solar Telescope}
\newacro{THEMIS}{T\'{e}lescope H\'{e}liographique pour l'Etude du 
   Magn\'{e}tisme et des Instabilit\'{e} Solaires}     %RR wow
\newacro{TR}{transition region}
\newacro{TRACE}{Transition Region and Coronal Explorer}
\newacro{TSI}{total solar irradiance}
\newacro{UT}{Universal Time}
\newacro{UV}{ultraviolet}
\newacro{VAULT}{Very high angular resolution ultraviolet telescope}
\newacro{VIRGO}{Variability of solar IRradiance and Gravity Oscillations}
\newacro{VTT}{Vacuum Tower Telescope}    %RR "vauteetee"
\newacro{XRT}{X-Ray Telescope}

\def\acp#1{\pdftooltip{\acs{#1}}{\acl{#1}}}  %% pdfcomment.sty Apr 26 2015 

\def\nl{,\ } %%\def\nl{\newline}  %% redefine as \newline for mail addresses

\def\ITA{Institute of Theoretical Astrophysics\nl
         University of Oslo\nl
         P.O. Box 1029, Blindern\nl N-0315 Oslo\nl Norway}

\def\LA{Lingezicht Astrophysics\nl 't Oosteneind 9\nl 4158\,CA Deil\nl 
        The Netherlands}

\long\def\startignore #1\stopignore{}   %% use \startignore....\stopignore

\def\rmit#1{{\it #1}}              %% italics (RR style, Kluwer)
\def\etal{\rmit{et al.}}           %% use \etal\ for space behind it        
           
\def\ie{\rmit{i.e.,}}              %% , required American (Webster 1681)
\def\eg{\rmit{e.g.,}}              %% , required American (Webster 1681)
\def\cf{cf.}                       %% no Latin, always Roman (Webster 1686)

\def\specchar#1{\uppercase{#1}}    %% redefine for A&A, small caps
\def\specand{ and }                %% eg H and K for ApJ
\def\specand{\,\&\,}               %% eg H&K for A&A
\def\CI{\mbox{C\,\specchar{i}}} 
\def\CII{\mbox{C\,\specchar{ii}}}

\def\CV{\mbox{C\,\specchar{v}}} 
 
\def\CaII{\mbox{Ca\,\specchar{ii}}}

\def\FeI{\mbox{Fe\,\specchar{i}}} 
\def\FeII{\mbox{Fe\,\specchar{ii}}}

\def\HI{\mbox{H\,\specchar{i}}} 
\def\HII{\mbox{H\,\specchar{ii}}} 
\def\Hmin{\hbox{\rmH$^{^{_-}}\!$}}      %% H^min, very elegant
\def\HeI{\mbox{He\,\specchar{i}}}

\def\MgI{\mbox{Mg\,\specchar{i}}} 
\def\MgII{\mbox{Mg\,\specchar{ii}}} 
\def\MgIII{\mbox{Mg\,\specchar{iii}}} 
 
\def\MnI{\mbox{Mn\,\specchar{i}}}

\def\NaI{\mbox{Na\,\specchar{i}}}
\def\NaII{\mbox{Na\,\specchar{ii}}}

\def\SiI{\mbox{Si\,\specchar{i}}} 
 
\def\SiIII{\mbox{Si\,\specchar{iii}}} 
\def\SiIV{\mbox{Si\,\specchar{iv}}}
\def\SiV{\mbox{Si\,\specchar{v}}}

\def\TiII{\mbox{Ti\,\specchar{ii}}}

\def\Halpha{\mbox{H\hspace{0.1ex}$\alpha$}} %% \Halpha\ for space behind it

\def\Hbeta{\mbox{H\hspace{0.2ex}$\beta$}}

\def\Lyalpha{\mbox{Ly$\hspace{0.2ex}\alpha$}}
\def\Lybeta{\mbox{Ly$\hspace{0.2ex}\beta$}}

\def\HeIDthree{\mbox{He\,\specchar{i}\,\,D$_3$}}

\def\NaID{\mbox{Na\,\specchar{i}\,\,D}}
\def\NaIDone{\mbox{Na\,\specchar{i}\,\,D$_1$}}

\def\MgIb{\mbox{Mg\,\specchar{i}\,b}}

\def\MgIbtwo{\mbox{Mg\,\specchar{i}\,b$_2$}}

\def\CaIIK{\mbox{Ca\,\specchar{ii}\,\,K}}       %% use \CaIIK\ for space
\def\CaIIH{\mbox{Ca\,\specchar{ii}\,\,H}}

\def\HK{\mbox{H{\specand}K}}
\def\KtwoV{\mbox{K$_{2V}$}}

\def\CaIR{\mbox{Ca\,\specchar{ii}\,8542\,\AA}} 

\def\MgIIk{\mbox{Mg\,\specchar{ii}\,\,k}}

\def\MgIIhk{\mbox{Mg\,\specchar{ii}{\specand}k}}
\def\hk{\mbox{h{\specand}k}}

\def\level #1 #2#3#4{$#1 \; ^{#2} \mbox{#3} ^{#4}$}   

 \def\rmD{{\rm D}} 
\def\rme{{\rm e}} \def\rmE{{\rm E}}

 \def\rmH{{\rm H}}

\def\kms{\hbox{km$\;$s$^{-1}$}}

\def\is{\!=\!}                             %% tighter spacing
\def\={\hbox{$\!=\!$}}                     %% no space around =
\def\ep{\:{\rm e}^}                        %% e^ with space and roman e
\def\rmit#1{#1}                 %% A&A & ApJ: latin abbreviations in Roman
\def\specchar#1{{\textsc{#1}}}  %% small caps for A&A
\begin{document}

%%\title{Ellerman bomb visibilities per LTE}
%%\title{Ellerman bomb visibilities per Saha-Boltzmann extinction estimate}
\title{H-alpha features with hot onsets.  I. Ellerman bombs}
\subtitle{}

\author{R. J. Rutten}

\institute{\LA \and \ITA} 

\date{Received May 7, 2015 /  Accepted January 13, 2016}

\abstract{Ellerman bombs are transient brightenings of the wings of
the Balmer lines that uniquely mark reconnection in the solar
photosphere.
They are also bright in strong \CaII\ and ultraviolet lines and in
ultraviolet continua, but they are not visible in the optical
continuum and the \NaID\ and \MgIb\ lines.
These discordant visibilities invalidate all published Ellerman bomb
modeling.
I argue that the assumption of Saha-Boltzmann lower-level populations
is informative to estimate bomb-onset opacities for these diverse
diagnostics, even and especially for \Halpha, and employ such
estimates to gauge the visibilities of Ellerman bomb onsets in all of
them.  
They constrain Ellerman bomb formation to temperatures
10\,000--20\,000\,K and hydrogen densities around 10$^{15}$ cm$^{-3}$.
Similar arguments likely hold for \Halpha\ visibility in other
transient phenomena with hot and dense onsets.}

\keywords{Sun: activity -- Sun: atmosphere -- Sun: magnetic fields}

\maketitle

%%%%%%%%%%%%%%%%%%%%%%%%%%%%%%%%%%%%%%%%%%%%%%%%%%%%%%%%%%%%%%%%%%%%%%%%%%%%
\section{Introduction}\label{sec:introduction}
%%%%%%%%%%%%%%%%%%%%%%%%%%%%%%%%%%%%%%%%%%%%%%%%%%%%%%%%%%%%%%%%%%%%%%%%%%%%
\citetads{1917ApJ....46..298E} 
discovered intense short-lived brightenings of the extended wings of
the Balmer \Halpha\ line at 6563\,\AA\ which are called ``Ellerman
bombs'' (henceforth EB) since
\citetads{1960PNAS...46..165M} 
or ``moustaches'' after their rediscovery by 
\citetads{1956Obs....76..241S}. 
The extensive \acp{EB} literature was reviewed by
\citetads{2013JPhCS.440a2007R}. 
Recent studies are discussed in
\citetads{2015ApJ...812...11V}. 
This paper treats \acp{EB} visibilities.

Below I refer frequently to a series of observational
\acp{EB} studies using high-quality imaging spectroscopy and
polarimetry with the \acl{CRISP} (CRISP,
\citeads{2008ApJ...689L..69S}) 
at the \acl{SST} (SST,
\citeads{2003SPIE.4853..341S}). 
Paper~I (\citeads{2011ApJ...736...71W}) 
established that \acp{EB}s are a purely photospheric phenomenon.
Paper~II (\citeads{2013ApJ...774...32V}) 
added evidence that \acp{EB}s mark magnetic reconnection of strong
opposite-polarity field concentrations and
treated their appearance in ultraviolet continuum images from the
\acl{AIA} (AIA, \citeads{2012SoPh..275...17L}) 
onboard the \acl{SDO}
(SDO, \citeads{2012SoPh..275....3P}). 
Paper~III
(\citeads{2015ApJ...812...11V}) 
presented the marked appearance of \acp{EB}s in ultraviolet \SiIV,
\CII\ and \MgII\ lines in spectra taken with the \acl{IRIS} (IRIS,
\citeads{2014SoPh..289.2733D}). 
Paper~IV (\citeads{2015ApJ...808..133R}) 
confirmed Ellerman's statement that \acp{EB}s are not obvious in the
optical continuum nor in the \NaID\ and \MgIb\ lines.

%% -------------------------------------------- Papers I-IV definition macros
%% add (Paper I) to the reference
\bibnote{2011ApJ...736...71W}{(Paper~I)}
%% define \PaperI as ADS clicker
\def\PaperI{\href{http://adsabs.harvard.edu/abs/2011ApJ...736...71W}
  {Paper~I}}
%% add (Paper II) to the reference
\bibnote{2013ApJ...774...32V}{(Paper~II)}
%% define \PaperII as ADS clicker
\def\PaperII{\href{http://adsabs.harvard.edu/abs/2013ApJ...774...32V}{Paper~II}}
%% add (Paper III) to the reference
\bibnote{2015ApJ...812...11V}{(Paper~III)}
%% define \PaperIII as ADS clicker 
\def\PaperIII{\href{http://adsabs.harvard.edu/abs/2015ApJ...812...11V}{Paper~III}}
%% add (Paper IV) to the reference
\bibnote{2015ApJ...808..133R}{(Paper~IV)}
%% define \PaperIV as ADS clicker 
\def\PaperIV{\href{http://adsabs.harvard.edu/abs/2015ApJ...808..133R}{Paper~IV}}
%%-------------------------------------------------------

In summary, the visibilities of \acp{EB}s are now well documented. 
They present a rich set of constraints with large and puzzling
variety.
\acp{EB}s appear very bright and with extraordinary spectral
extent in Balmer-line wings, also bright but less extended in the
wings of \CaII\ \HK\ and \CaIR, very bright in the \acp{IRIS} \CII\
1334 \& 1335\,\AA\ and \SiIV\ 1394 \& 1403\,\AA\ doublets,
and bright and very bright in \acp{AIA}'s 1700 and 1600\,\AA\
channels, respectively. 
However, they are transparent in the optical continuum and also absent
or nearly absent in optical neutral-atom lines (\FeI, \NaI, \MgI).

These diverse \acp{EB} visibilities and non-visibilities invalidate
the \acp{EB} modeling published so far.
\citetads{1983SoPh...87..135K}, 
\citetads{2010MmSAI..81..646B}, 
\citetads{2013A&A...557A.102B} 
and
\citetads{2014A&A...567A.110B} 
performed \Halpha\ synthesis from best-fit ad-hoc perturbations of a
static 1D standard model of the solar atmosphere to reproduce observed
\Halpha\ wing brightenings. 
These modeling efforts were summarized in \PaperIV\  together with
their failures: non-reproduction of the observed photospheric 
anchoring of \acp{EB}s in internetwork lanes (\PaperI),
non-reproduction of \acp{EB} brightening in \acp{IRIS} \SiIV\ and
\CII\ lines (\PaperIII), and non-reproduction of \acp{EB} non-visibility
in the \NaID\ and \MgIb\ lines (\PaperIV).

In addition to these ad-hoc \Halpha-synthesis attempts there have been
numerical \acp{MHD} simulations of \acp{EB} reconnection by
\citetads{2009A&A...508.1469A}, 
but without any spectral synthesis or verification, and by
\citetads{2013ApJ...779..125N} 
who probably did not emulate an \acp{EB} nor observed one (\PaperIII).

In the present paper I discuss \acp{EB} visibilities
assuming \acdef{LTE} for line extinctions, \ie\
Saha-Boltzmann lower level populations.
This assumption may seem surprising for such utterly non-equilibrium
fast-changing dynamic phenomena, since \acp{EB}s mark magnetic
reconnection at small spatial and short temporal scales (\PaperI,
\PaperII), including near-instantaneous heating, bi-modal jet
formation (\PaperI, \PaperIII), hot-cloud expansion (\PaperIII), and
possibly cooling clouds in their aftermath (\PaperIII, \PaperIV). 

Assuming \acp{LTE} (\ie\ validity of Boltzmann and Saha partitioning
over atomic levels and ionization stages) seems totally out of the
question for \acp{EB}s.  
Even the assumption of statistical equilibrium (SE, time-independent
level and stage populations) seems highly questionable.
This assumption is the basis for both non-LTE modeling (NLTE, meaning
\acp{SE} including all pertinent bound-bound and bound-free population
processes) and coronal-equilibrium modeling (CE, admitting only
collisional excitation and ionization and only radiative deexcitation
and recombination including dielectronic), but \acp{SE} is unlikely to
hold in \acp{EB}s.
These then require non-equilibrium modeling (non-E, \ie\ solving all
pertinent population and radiation equations time-dependently).

My \acp{LTE} approach below (Sect.~\ref{sec:analysis}) may seem even
more surprising in view of my background (\eg\
\citeads{2003rtsa.book.....R}, henceforth RTSA). 
%% add (RTSA) to the reference
\bibnote{2003rtsa.book.....R}{(RTSA)}
%% define \PaperI as ADS clicker
\def\RTSA{\href{http://adsabs.harvard.edu/abs/2003rtsa.book.....R}
{RTSA}} Let me emphasize at the outset that realistic \acp{EB}
modeling does require 3D time-dependent non-E \acp{MHD} simulation
including non-E 3D spectral synthesis. 
The actual \acp{NLTE} departures may reach $10^{10}$ (10~dex) or more,
far beyond the 0.01--0.1 dex departures discussed in
abundance analyses.
However, such computations have been accomplished in full only
for 1D cases, in 2D only with questionable shortcut recipes. 
The 3D case remains prohibitive
(\citeads{2012A&A...539A..39C}). 
Leenaarts \etal\
(\citeyearads{2012ApJ...749..136L}, 
\citeyearads{2015ApJ...802..136L}) 
therefore restricted their numerical 3D \Halpha\ formation studies to
a single simulation snapshot.

In this paper I first argue on the basis of published modeling that
the \acp{LTE} simplification can serve to gauge potential \acp{EB}
visibilities. 
The crux is to assume \acp{LTE} not all the time but only during hot
and dense \acp{EB} onsets 
and only for line extinctions, not for line source functions.
I then combine this onset assumption with the observed \acp{EB}
visibilities to define the probable \acp{EB} parameter domain.

The organization of the paper is as follows.
The next section gives background.
It starts with basic equations and their implementation. 
It then treats three
published solar-atmosphere models: the 1D SE static model of
\citetads{2008ApJS..175..229A}, 
the 2D time-dependent non-E MHD simulation of
\citetads{2007A&A...473..625L}, 
and the 1D SE static model of
\citetads{2009ApJ...707..482F}. 
All three were intended to mimic the solar atmosphere, but I treat
them as fictitious stellar atmospheres named ALC7, HION and FCHHT-B.
The HION atmosphere inspired this study.
The ALC7 and FCHHT-B chromospheres supply instructive line formation
demonstrations even though they are poor renderings of the actual
solar chromosphere.
From these examples I distill recipes for estimating line extinction
(Sect.~\ref{sec:recipes}).
Sect.~\ref{sec:terminology} concludes the background part by defining
observational terms to avoid confusions.

The analysis (Sect.~\ref{sec:analysis}) boils down to plotting and
interpreting classical diagrams.
Figure~\ref{fig:CE-LTE} compares \acp{LTE} to \acp{CE} ionization
equilibria, the latter following \citetads{1969MNRAS.142..501J},
because within the \acp{SE} assumption these represent extremes.
Figure~\ref{fig:extLTE} emulates the famous diagram\footnote{ I taught
making Saha-Boltzmann graphs to hundreds of students at Utrecht and
elsewhere with a lab exercise ``Cecilia Payne'' (available on my
website). 
It uses a fictitious and unpronounceable didactic element called
``Schadeenium'' after Utrecht astrophysicist Aert Schadee
(1936--1999), who invented it for teaching in the 1970s.} 
in Fig.~8 of \citetads{1925PhDT.........1P} (first printed in Fig.~1
of \citeads{1924HarCi.256....1P} and redrawn in Fig.~5-6 of
\citeads{1973itsa.book.....N}). 

Sects.~\ref{sec:discussion} and \ref{sec:conclusion} add
a brief discussion and conclusions.

%%%%%%%%%%%%%%%%%%%%%%%%%%%%%%%%%%%%%%%%%%%%%%%%%%%%%%%%%%%%%%%%%%%%%%%%%%%%
\section{Background} \label{sec:background}
%%%%%%%%%%%%%%%%%%%%%%%%%%%%%%%%%%%%%%%%%%%%%%%%%%%%%%%%%%%%%%%%%%%%%%%%%%%%

\subsection{Equations and implementation} \label{sec:equations}
%%%%%%%%%%%%%%%%%%%%%%%%%%%%%%%%%%%%%%%%%%%%%%%%%%%%%%%%%%%%%%%%%%
This section specifies the basic line formation equations
used to compute and interpret Figs.~~\ref{fig:ALC7} and 
\ref{fig:FCHHT-B} -- \ref{fig:extLTE} below.

The line extinction per cm path length is given by (\RTSA\ Eq.\ 9.6):
\begin{equation}
      \alpha^l
       =   \frac{\pi e^2}{m_\rme c} \, 
           \frac{\lambda^2}{c}\,
            b_l \, \frac{n_l^{\rm LTE}}{N_\rmE} \, N_\rmH \, A_\rmE
            \, f_{lu} \, \varphi
            \left[1-\frac{b_u}{b_l} \,
                    \frac{\chi}{\varphi}\,
                    \ep{-hc/\lambda kT}\right],
        \label{eq:ext}
\end{equation} 
where $e$ is the electron charge, $m_\rme$ the electron mass, $c$ the
velocity of light, $\lambda$ the line wavelength, $b_l$ the
lower-level and $b_u$ the upper-level \acp{NLTE} population departure
coefficient defined as $b \equiv n/n^{\rm LTE}$,
$n_l^{\rm LTE}/N_\rmE$ the lower-level population density given by the
Saha and Boltzmann laws as fraction of the total element density
$N_\rmE$, $N_\rmH$ the total hydrogen density, $A_\rmE$ the elemental
abundance, $N_\rmE/N_\rmH$, $f_{lu}$ the lower-to-upper oscillator
strength, $\varphi$ the area-normalized extinction profile in
wavelength, $\chi$ the induced-emission profile, $h$ the Planck
constant, $k$ the Boltzmann constant, and $T$ the temperature.
The profiles $\varphi$ and $\chi$ differ in the case of partial
frequency redistribution (PRD) but are equal for complete
redistribution (CRD).

The line source function is given by:
\begin{equation}
S^l = (1-\varepsilon-\eta)\,J + \varepsilon\,B + \eta\,S^\rmD,
\label{eq:Sline}
\end{equation}
where $\varepsilon$ is the photon destruction probability (the
fraction of photoexcitations followed by direct collisional
upper-to-lower deexcitation), $\eta$ the photon conversion probability
(the fraction of photoexcitations followed by detour upper-to-lower
paths), $J$ the angle-averaged intensity (profile-averaged
$\overline{J}$ for \acp{CRD} or a wavelength-dependent mix of the
monochromatic $J_\lambda$ for \acp{PRD}), $B$ the Planck function for
the local temperature, and $S^\rmD$ a formal source function which
collectively describes all multi-level upper-to-lower detour chains
(sometimes written as $B(T^\rmD)$ with $T^\rmD$ a formal detour
temperature as in Eqs.~8.5--8.8\footnote{Correction: delete the
minus in the second version of Eq.~8.8.} of
\citeads{1968slf..book.....J}). 
Eq.~\ref{eq:Sline} is the multi-level extension of the two-level
version including scattering (Eq.~10 of
\citealp{Hummer+Rybicki1967}, %% Comp Phys not in ADS @ add!!
Eq.~1.95 of \citeads{1979rpa..book.....R}, 
Eqs.~3.94 and 3.105 of \RTSA, Eq.~14.34 of
\citeads{2014tsa..book.....H}) 
and follows from that following Sect.~8.1 of
\citetads{1968slf..book.....J}. 

For the Wien approximation with $hc/\lambda kT \!>\!1$ the simple
expressions hold:
\begin{eqnarray}
  \alpha^l \!&\approx&\! b_l \,\, \alpha^{\rm LTE}\,  
   \label{eq:extWien} ,\\
  S^l \!&\approx&\! (b_u/b_l)\,\, B \, 
   \label{eq:SlineWien} .
\end{eqnarray}
This is the case for all diagnostics treated here (\CaIR\ reaches
$\lambda kT \is hc$ at 16\,800\,K, \Halpha\ at 21\,900\,K).
These equations illustrate that extinction \acp{NLTE} 
($b_l\!\neq\!1$) and source function \acp{NLTE} ($b_u/b_l\!\neq\! 1$)
are different entities, as are extinction and source function themselves.

The emergent intensity which conveys our telescopic diagnostics is
usually well approximated by the Eddington-Barbier estimate
$I \approx S(\tau\is1)$ for optically thick formation with $\tau$ the
optical depth along the line of sight. 
For non-irradiated optically thin slabs it is
$I \approx j\,D = \tau\,S$, with emissivity $j=\alpha\,S$ per cm,
geometrical thickness $D$, and optical thickness $\tau$.

In the thick case extinction \acp{NLTE} affects only the
representative $\tau\is1$ sampling depth (Eq.~\ref{eq:extWien}),
whereas source function \acp{NLTE} affects the intensity directly.
In the thin case both extinction and source function \acp{NLTE}
affect the intensity directly by together setting the emissivity.

\acp{NLTE} \acp{SE} spectrum synthesis codes, such as the workhorse RH code of
\citetads{2001ApJ...557..389U}, 
solve the coupled population and radiation equations for all
transitions and locations pertinent to selected lines, or even the
whole spectrum. 
I use RH in the latter mode in Sect.~\ref{sec:SE} below.

%% fig:ALC7
%===========================================================================
\begin{figure*}
  \sidecaption
  \includegraphics[width=120mm]{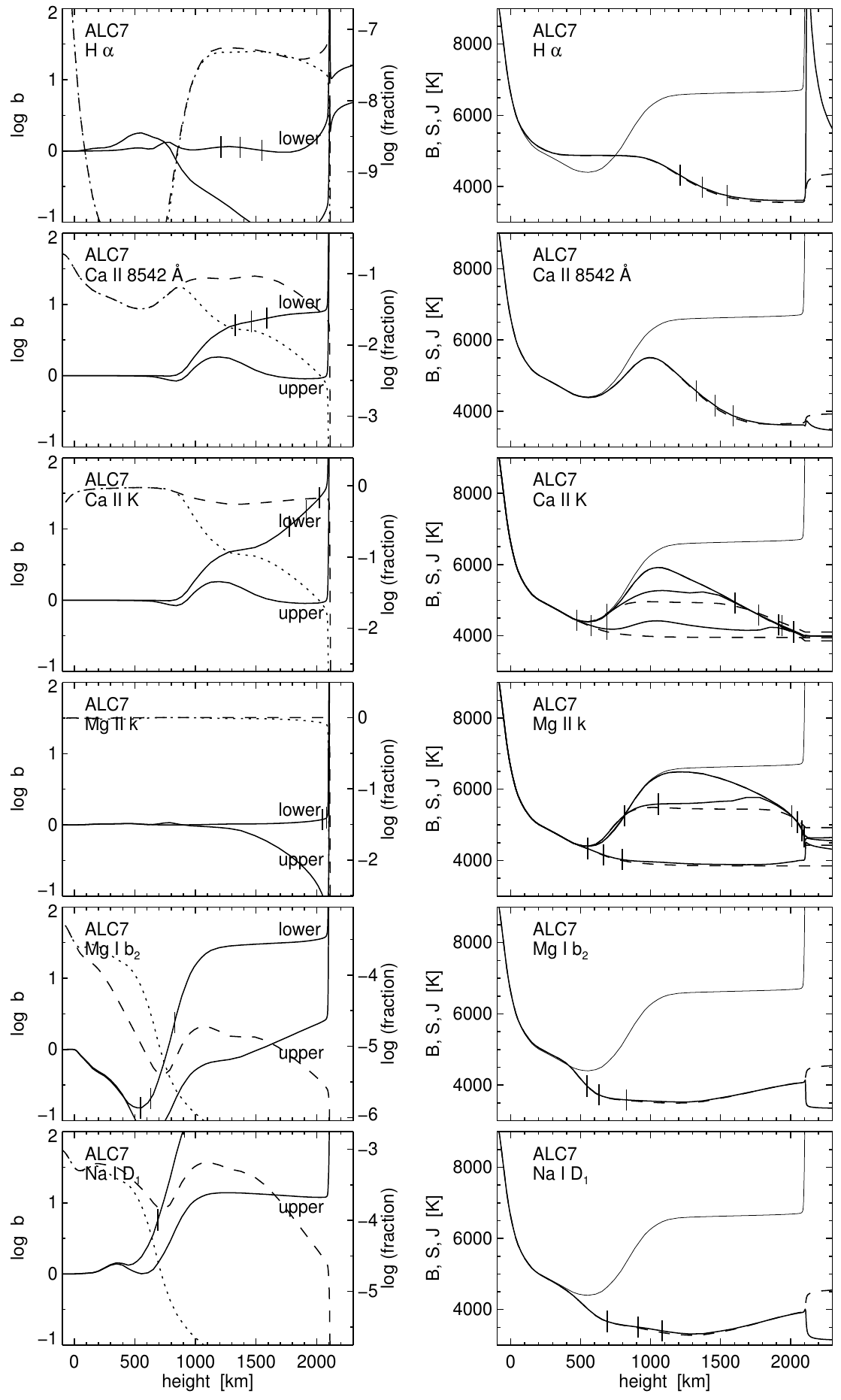}
  \caption[]{\label{fig:ALC7} 
  \acp{SE} formation of \acp{EB} diagnostics in the ALC7 model
  atmosphere of \citetads{2008ApJS..175..229A} 
  computed with the RH code of
  \citetads{2001ApJ...557..389U}.\\ 
  {\em Left, solid\/}: logarithmic \acp{NLTE} population
  departure coefficients $b\equiv n/n^{\rm LTE}$ as function of
  height, respectively $b_l$ for the lower level and $b_u$ for the
  upper level of the line identified in each panel. \\
    {\em Left, dashed\/}: logarithmic lower-level populations
  normalized by the total element density (axis scales to the right).\\
  {\em Left, dotted\/}: the same fractional population in \acp{LTE}.\\
  {\em Right\/}: Planck function $B_\lambda$ ({\em thin solid\/}, the
  same in all panels), angle-averaged intensity $\overline{J}$ or
  $J_\lambda$ ({\em dashed\/}), line source function $S_\lambda$ ({\em
  solid\/}). 
  These quantities are plotted as representative temperatures to avoid
  the Wien sensitivity of the Planck function and so obtain directly
  comparable scales.
  For \acp{PRD} lines (\CaIIK\ and \MgIIk) $S_\lambda$ and $J_\lambda$
  are shown for line center and the emergent-profile peaks and dips.
  The three vertical ticks define the heights where total optical
  depth $\tau_\lambda \is 3, 1, 0.3$, respectively (in the lefthand
  column only for line center on the $b_l$ curve).
  The $S \!<\! B$ splits correspond to the $b_u\!<\! b_l$ divergences
  at left.\\
  {\em Remarkable\/}: at left the very deep dip in the dashed \Halpha\
  population curve, the relative flatness of the \CaIR\ and \CaIIK\
  population curves, the utter flatness of the \MgIIk\ population
  curve, and the large chromospheric overpopulations from photon
  suction for the \MgIbtwo\ and \NaIDone\ lower levels.
  At right the discordantly high value of $S \!\approx\! J$ for
  \Halpha\ across the temperature minimum from backscattering and the
  overall similarity of the scattering $S \!\approx\! J$ declines of
  all lines, including independently scattering \acp{PRD} wings.  
  \vspace{2ex}\mbox{}} 
\end{figure*}
%===========================================================================

For the simpler \acp{LTE} extinction evaluations I updated \acp{IDL}
codes to compute ionization and molecule mixes and partial pressures
for given elemental composition that were developed by J.~S\'anchez
Almeida (\citeyearads{1992SoPh..137....1S}, 
\citeyearads{1997ApJ...491..993S}) 
and are partly based on
\citetads{1974SoPh...35...11W} 
following \citet{Mihalas1967}. 
I found them in the
\href{https://github.com/aasensio/lte/tree/master/idl}{github LTE
repository} of A.~Asensio Ramos and extended them with data and
routines in the SolarSoft CHIANTI package (\eg\
\citeads{1997A&AS..125..149D}, 
\citeads{2013ApJ...763...86L}). 
My codes are available under IDL on
\href{http://www.staff.science.uu.nl/~rutte101}{my website}.

\subsection{ALC7: EB diagnostics in 1D SE modeling} 
\label{sec:SE}
%%%%%%%%%%%%%%%%%%%%%%%%%%%%%%%%%%%%%%%%%%%%%%%%%%%%%%%%%%%%%%
This section presents and discusses the formation of \acp{EB}
diagnostics in quiescent conditions in order to establish their normal
behavior.

Figure~\ref{fig:ALC7} 
presents \acp{EB} diagnostics assuming \acp{SE} and E.H.~Avrett's
latest empirical static 1D model
(\citeads{2008ApJS..175..229A}). 
Below I question the validity of such plane-parallel models, but they
do remain exemplary for demonstrating fully understandable solar-like
line formation, in this case in
the computational atmosphere ``ALC7''.

Figure~\ref{fig:ALC7} shows extinction parameters at left, source
function parameters at right.
They were computed with the 1D RH version of
\citetads{2001ApJ...557..389U} 
including \acp{NLTE} line blanketing with the full compilation
of \citetads{2009AIPC.1171...43K}. 
\acp{PRD} was adopted for \CaIIK\ and \MgIIk; \acp{CRD} for the
others.
The \acp{IRIS} \CII\ and \SiIV\ lines were excluded because they form
in the ALC7 \acp{TR} and pose numerical difficulties.

In the lefthand column the divergences between the dashed and dotted
fractional population curves correspond to the departure of $\log b_l$
from zero.
All plots show increasing $b_u < b_l$ divergence with height due
to photon losses.
Equation \ref{eq:SlineWien} implies that these
divergences translate into equal divergences between
$S$ and $B$, but the equivalent temperature
representation used at right produces larger divergence at longer
wavelength for given $b_u/b_l$. 

\Halpha\ (top panels of Fig.~\ref{fig:ALC7}) has discordant formation
across the upper ALC7 photosphere because this is transparent in this
high-excitation line (dashed curve in the first panel). 
The radiation field there builds up from backscattering from the
opaque overlying ALC7 chromosphere (\citeads{2012A&A...540A..86R}). 
The lower-level population, hence the \Halpha\ line extinction, nearly
equals the Saha-Boltzmann value up to the ALC7 \acp{TR} thanks to the
huge \Lyalpha\ opacity
producing $S^l \!\approx\! J \!\approx\! B$ in this line, 
with small wiggles from radiative \Lyalpha\ smoothing
(Sect.~\ref{sec:spatial}). 

The \Halpha\ source function (second panel) behaves like a standard
scattering one, \ie\ domination of the first two terms in
Eq.~\ref{eq:Sline} that together cause a scattering decline in
$b_u/b_l$.
\Halpha\ has often been called ``photoelectric'' following
\citetads{1957ApJ...125..260T}, meaning dominance of the third term in
Eq.~\ref{eq:Sline} over the second, but across the ALC7 chromosphere
it is simply a resonance-scattering line with a source function
decline like all others in Fig.~\ref{fig:ALC7}

Since the ALC7 chromosphere is near-isothermal there is good
resemblance to the classic demonstration of two-level scattering
by \citetads{1965SAOSR.174..101A}, 
except that $\varepsilon$ is not constant. 
 
The outward $S$ decline from such scattering (``$\sqrt{\varepsilon}$
law'') causes a dark line core which is often called
``self-absorption'' in ultraviolet spectroscopy. 

In the ALC7 atmosphere the \Halpha\ core forms halfway the
chromosphere, but most escaping photons actually
originated in the deep ALC7 photosphere where $\varepsilon$
reaches unity.
For \Halpha\ the ALC7 chromosphere is only a scattering
re-director. 
If it were absent the emergent \Halpha\ profile would remain the same,
originating in a similar scattering decline within the
photosphere. 

In actual solar \Halpha\ images such scattering (but 3D)
obliterates the granulation imprint although this has larger
intrinsic contrast than chromospheric fine structure, making the
latter visible.
Where the chromospheric slab has larger density the
$\tau\is1$ location moves outward along the scattering decline, making
\Halpha\ core darkness primarily a density diagnostic
(\citeads{2012ApJ...749..136L}). 

\CaIR\ (second row) forms at about the same height as \Halpha\ in
ALC7, but has a better-behaved source function that follows the
temperature until it drops down from scattering.
The $b_l$ curve in the lefthand panel shows substantial increase from
photon losses in the \CaII\ infrared lines. 
The initial dip in the population curve follows the temperature
through the Boltzmann sensitivity.  

\CaIIK\ behaves very similar to \CaIR, but its core forms higher and
\acp{PRD} applies, shown by sampling $S$ and $J$ at the K$_3$, K$_2$
and K$_1$ wavelengths (\cf\
\citeads{1975ApJ...199..724S}). 
The $b_l$ curve first rises from photon losses in the \CaII\ infrared
lines that are shared through collisional lower-level coupling,
then from photon losses in \HK.
They offset the \CaII\ depletion by ionization that would
occur in \acp{LTE} (dotted curve).

\MgIIk\ behaves much as \CaIIK, but the 18 times larger Mg abundance
makes it form in the ALC7 \acp{TR}.
Its \acp{PRD} wings form across the ALC7 chromosphere.  
The higher ionization energy (15.0\,eV instead of 11.9\,eV) makes
\MgII\ fully dominant throughout the ALC7 chromosphere (dashed curve
in the lefthand panel), resulting in pure \acp{LTE} extinction
and textbook decline of the $b_u$ curve.

\MgIbtwo\ and \NaIDone\ (last rows) have 
their $\tau\is1$ escape levels in the onset of the ALC7 chromosphere,
but the escaping photons originate so deep that the line core
intensity does not sense the ALC7 chromosphere.
These lines have large extinction increase over \acp{LTE} from
minority-stage photon suction 
(\citeads{1992A&A...265..237B}). 
Below that, \MgI\ shows a steep departure-coefficient decline from
overionization by ultraviolet radiation from below
(\citeads{2012A&A...540A..86R}). 

In summary, Fig.~\ref{fig:ALC7} shows that \acp{LTE} extinction is a
reasonable assumption for these \acp{EB} diagnostics throughout the
ALC7 chromosphere. 
In particular, it is valid for the Balmer lines and the Balmer
continuum and for \MgIIk, whereas it even represents
considerable underestimation for the \CaII\ lines and \NaIDone.
It is an overestimation only for \MgIbtwo.

\acp{EB}-like hot and dense features embedded in such an atmosphere
will have larger collisional coupling, especially when hydrogen
ionization boosts the electron density and $\varepsilon$, and be yet
closer to \acp{LTE}.

%% fig:HION
%===========================================================================
\begin{figure}
%%  \centerline{\includegraphics[width=\columnwidth]{\deffigs/fig-HION}}
  \centerline{\includegraphics[width=\columnwidth]{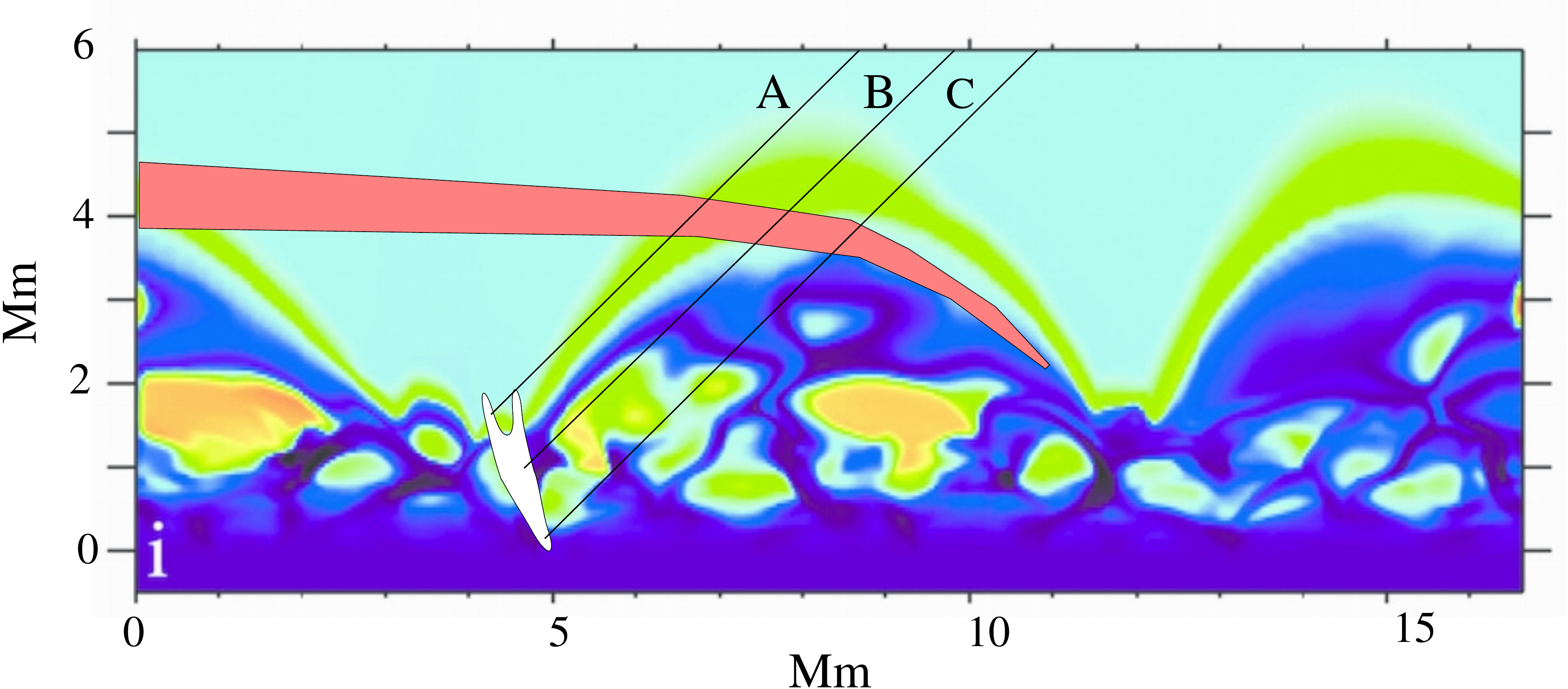}}
  \caption[]{\label{fig:HION} 
  Schematic \acp{EB} in the HION atmosphere. 
  A sketched \acp{EB} (white, near $x\is5$) is superimposed on the
  last panel of Fig.~1 of
  \citetads{2007A&A...473..625L} 
  in which the black--blue--green--orange color coding quantifies
  \acp{NLTE} overpopulation of the \HI\ $n\is2$ level ranging from
  1 to 10$^{14}$. 
  The superimposed pink arch represents a schematic \Halpha\ fibril.
  Slanted lines of sight ($\mu=0.71$) to the top, middle, and bottom
  of the \acp{EB} are marked A, B, and C.
  }
\end{figure}
%===========================================================================

%% fig:FCHHT-B
%===========================================================================
\begin{figure*}
  \sidecaption
  \includegraphics[width=120mm]{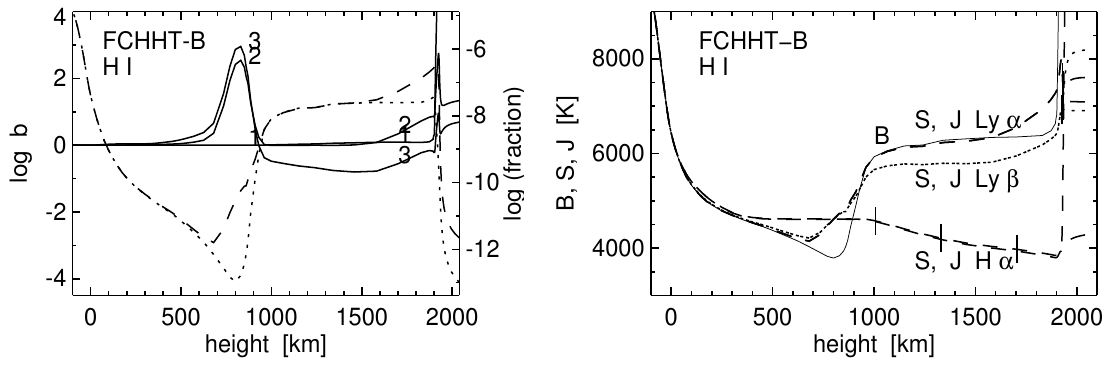}
   \caption[]{\label{fig:FCHHT-B} 
  \acp{SE} formation of \Halpha, \Lyalpha, and \Lybeta\ in the FCHHT-B
  atmosphere of \citetads{2009ApJ...707..482F} 
  in the format of Fig.~\ref{fig:ALC7}, using RH results from
  \citetads{2012A&A...540A..86R}. 
  The lefthand panel shows \acp{NLTE} departure coefficients $b$ for
  hydrogen levels $n\is1,2,3$.
  The dashed and dotted \acp{NLTE} and \acp{LTE} fractional population
  curves are for $n\is2$.
   The righthand panel shows the corresponding line source functions.
 \vspace{2ex}\mbox{}} 
\end{figure*}
%===========================================================================

\subsection{HION: \Halpha\ extinction boost to past hot LTE}
\label{sec:temporal}
%%%%%%%%%%%%%%%%%%%%%%%%%%%%%%%%%%%%%%%%%%%%%%%%%%%%%%%%%%%%%%%%
This section presents the motivation for assuming \acp{LTE}
extinction in hot onsets of dynamical phenomena.

This assumption was inspired by the numerical non-E \acp{MHD} simulation of
\citetads{2007A&A...473..625L}. 
It did not address \acp{EB}s but a vertical plane containing
two opposite-polarity strong-field concentrations resembling solar
network with less magnetic internetwork in between.
Its theme was \acp{non-E} modeling by permitting and evaluating
time-dependent hydrogen ionization and recombination. 
By being 2D and \acp{MHD} it represented a sequel to the seminal 1D
non-E \acp{HD} simulation of
\citetads{2002ApJ...572..626C} 
in which the physics of \acp{non-E} hydrogen population processes was
analyzed for acoustic shocks in the lower solar atmosphere. 

Below I refer so much to this simulation 
that I treat it as the atmosphere of a hypothetical star called
``HION'' (for \HI\ ionization).  
Its atmosphere exists only computationally but yet has solar-like properties.

In the HION atmosphere shocks occur copiously as field-guided ones
producing dynamic fibrils near the magnetic concentrations and as less
field-restricted ones in the quieter internetwork 
that resemble the acoustic 1D shocks of
\citetads{1997ApJ...481..500C}.

Figure~\ref{fig:HION} superimposes a sketch of an \acp{EB}
based on the \acp{SST} images in Papers I-IV on a snapshot of
the HION atmosphere.
I use the latter as a scenic illustration of how the actual
shock-ridden lower solar atmosphere around an \acp{EB} may look like.
The two opposite-polarity field concentrations at $x \is 4$ and
$x\is12$\,Mm closely resemble the actual kilogauss magnetic
concentrations that constitute solar network.
Besides them and elsewhere finely-structured thin shocks (black to
dark blue) run up and push the transition region (TR, onset of the
light-blue area) to large heights, of order 3~Mm.  
Large cool clouds appear behind the shocks (green to orange).
This scene is very dynamic: the HION \acp{TR} in HION dances up and
down over 2--4~\,Mm height in internetwork and over 1.5--2.5~Mm
in and near the field concentrations.

What does this HION scene have to do with assuming \acp{LTE}? 
(It actually depicts \Halpha\ \acp{NLTE} $b_l$ departures of twelve
orders of magnitude in the orange clouds!)
The essential \acp{non-E} hydrogen-ionization physics is that
collisional transitions between \HI\ $n\is1$ and $n\is2$ are so scarce
in cool post-shock gas that there hydrogen ionization/recombination
balancing is slow
(\citeads{1980A&A....87..229K}; 
\citeads{2002ApJ...572..626C}), 
leaving the ion state and with it the top of the hydrogen atom highly
overpopulated with respect to 
Saha-Boltzmann partitioning during the 3--5 minute cool intermezzo
until the next shock passes.

The shocked HION atmosphere is sampled in Fig.~2 of
\citetads{2007A&A...473..625L} 
with the bottom panels showing the \HI\ $n\is2$ population governing
the extinction of \Halpha\ (Eq.~\ref{eq:ext}).
The first key point is that it reaches the momentary \acp{LTE} value
in shocks.
This results from high temperature (about 7\,000\,K) and large
electron density increase from hydrogen ionization (above 1 percent),
to values $10^{10}-10^{11}$\,cm$^{-3}$,
together boosting $\varepsilon$ for \Lyalpha\ sufficiently to
couple $n\is2$ to the ground state within the shocks.
The second key point is that in between successive shocks the $n\is2$
population remains at or near this high level, even while the
temperature drops thousands of degrees, 
because the top of the hydrogen atom is decoupled from the ground
state during the retarded recombination.

Thus, both the \Halpha\ and Balmer-continuum opacities are near-LTE in
the shocks and remain at these high values between them.  
The gigantic overpopulations indicated by the orange clouds in
Fig.~\ref{fig:HION} and the corresponding large separations between
the thick and thin curves in the bottom panels of Fig.~2 of
\citetads{2007A&A...473..625L} 
simply mean that the actual $n\is2$ population doesn't track the cool
intermezzi.

\subsection{FCHHT-B: \Halpha\ extinction boost to nearby hot LTE}
\label{sec:spatial}
%%%%%%%%%%%%%%%%%%%%%%%%%%%%%%%%%%%%%%%%%%%%%%%%%%%%%%%%%%%%%%%%
This section addresses the effect of \Lyalpha\ scattering near
hot features.

Inter-shock temporal constancy of the $n=2$ population also implies
constancy of the \Lyalpha\ source function which has
$S\!\approx\!J\!\approx\!b_2 \, B(T)$ where hydrogen is
predominantly neutral (Eqs.~\ref{eq:Sline} and \ref{eq:SlineWien})
which it remains even in the HION shocks.
\citetads{2007A&A...473..625L} 
did not admit \Lyalpha\ or other Lyman 
radiation in HION, but it is interesting to note that, in addition to
the temporal \Halpha\ extinction smoothing to the hottest preceding
instances in the near past due to slow recombination settling, there
is also spatial \Halpha\ extinction smoothing to nearby hot instances.
This is due to \Lyalpha\ resonant scattering.
I illustrate it in Fig.~\ref{fig:FCHHT-B} using the FCHHT-B atmosphere
of \citetads{2009ApJ...707..482F}. 

The FCHHT-B atmosphere is similar in construction and properties to
the ALC7 atmosphere. 
The main difference is that ALC7 obtains chromospheric extent from
adding turbulent pressure constrained by observed non-thermal line
widths while FCHHT-B obtains extent from best-fit manipulation of the
local gravity. 
The FCHHT-B atmosphere has a low high-lying temperature minimum
adjacent to an abrupt rise to a near-isothermal chromosphere ($B$
curve in Fig.~\ref{fig:FCHHT-B}) intended to reproduce dark infrared
CO-line cores.

I use this sudden rise as emulating a hot feature embedded in a cool
atmosphere. 
The downward radiation from this hot edge may be seen as representing
how a cylindrical hot \acp{EB} irradiates cool gas around it.
The large $b_2$ and $b_3$ peaks in
Fig.~\ref{fig:FCHHT-B} arise from downward Lyman-line
scattering from the overlying chromosphere.
Even at the very large optical depths which the Lyman lines reach
there, their local scattering diffuses some of their intense
chromospheric $J \approx B$ radiation into the cool
underlying FCHHT-B layers.
The \Lyalpha\ irradiation boosts the $n\is2$ population and with it
the \Halpha\ extinction by a factor 400 at 100~km below the edge.

\Lybeta\ has a similar off-edge $S \approx J$ scattering halo as
\Lyalpha.
Over $h\is 1000-1500$~km the \Lyalpha\ radiation confinement causes
$b_2 \!\approx\! b_1 \!\approx\! 1$, so there \Lybeta\ has the same
degree of source function \acp{NLTE} as \Halpha\ since they share
$n\is3$ as their upper level (Eq.~\ref{eq:SlineWien}) and its
depopulation by \Halpha\ photon losses.
Their $S$ divergence from $B$ at right differs through the
representation as formal temperatures (which removes Planck function
sensitivity to wavelength).

\subsection{Recipes for  \Halpha\  extinction in EB  onsets}
\label{sec:recipes}
%%%%%%%%%%%%%%%%%%%%%%%%%%%%%%%%%%%%%%%%%%%%%%%%%%%%%%%%%%%%%%%%
I now apply the above to \acp{EB} onsets.
The upshot from Sects.~\ref{sec:temporal} and \ref{sec:spatial} is
that for hot dense instances in the low solar atmosphere one may:
(1) assume Saha-Boltzmann lower-level population for the extinction
coefficient of \Halpha, (2) apply such boosting also as an opacity halo to
cooler surrounding gas, and (3) maintain it during cooler subsequent
episodes.
These are the recipes applied in this paper.

Similar to the shocks pervading the HION atmosphere for which these
recipes hold, \acp{EB}s are momentary heating events in the low
atmosphere but yet deeper, hotter and denser.
I therefore suggest that the LTE recipes also apply to the
hydrogen $n\is 2$ population in \acp{EB}s.
They then hold for both \Halpha\ extinction and Balmer-continuum
extinction.

Do the recipes apply also to other \acp{EB} diagnostics?
This question is twofold: is Saha-Boltzmann population partitioning a
good approximation for their extinction at hot moments at high
density, and do they possess a memory for such momentarily high
extinction as the Balmer lines and continuum have?

In contrast to \Halpha, most other \acp{EB} diagnostics are
ground-state resonance lines and have extinction near or at the
\acp{LTE} value if the pertinent ionization stage is the dominant one.  
For them, in particular \SiIV, the question is whether Saha applies in
photospheric-density gas heated to high temperature.

The second recipe issue is the presence of a memory for previous
hot-feature extinction due to retarded recombination. 
This cannot be illustrated with static modeling. 
However, the \HI\ example shows that it occurs when there is a large
initial jump in the term structure of the lower stage
(\citeads{2002ApJ...572..626C}). 
\HeI\ is an obvious candidate
(\citeads{2014ApJ...784...30G}), 
but unfortunately there are no reports of \acp{EB} visibility in
\HeIDthree\ or 10\,830\,\AA.
The \NaID, \MgIb, \CaII\ and \MgII\ lines will be much less retarded
(\cf\ \citeads{2011A&A...528A...1W}; 
\citeads{2013ApJ...772...89L}). 
However, retarded recombination may well apply to \CII\ 1334 \&
1335\,\AA\ and \SiIV\ 1394 \& 1403\,\AA\ because their preceding ions
have large initial jumps: \CI\ has a 13\,eV gap between low and high
levels, the \SiIII\ resonance lines are like \Lyalpha\ near 1200\,\AA.
These large steps slow down recombination to the lower stage when
that becomes dominant in cool gas, just as for \HI\ in HION
post-shock gas.

\subsection{Terminology} \label{sec:terminology}
%%%%%%%%%%%%%%%%%%%%%%%%%%%%%%%%%%%%%%%%%%%%%%%%%%

This section defines observational nomenclature used in the
analysis in Sect.~\ref{sec:analysis}.

\paragraph{EB, MC, FAF, IB.}
%%%%%%%%%%%%%%%%%%%%%%%%
For clarity I list and define various \acp{EB}-related observed
phenomena here.

Ellerman bombs (EB) adhere to the definition by
\citetads{1917ApJ....46..298E}: 
sudden intense brightenings of the extended \Halpha\ wings
(``moustaches'') that occur exclusively in complex bipolar
active regions and have the diverse visibilities described in
Sect.~\ref{sec:introduction}.

Magnetic concentrations (MC) denote the kilogauss strong-field
elements that make up solar network. 
They are best known as faculae or G-band bright points, but also
appear as bright points in the 1700\,\AA\ continuum and especially in
the blue wing of \Halpha\
(\citeads{2006A&A...449.1209L}). 
The latter property has caused frequent confusion with \acp{EB}s
(\citeads{2013JPhCS.440a2007R}). 
\PaperI\ showed that at high resolution upright flame morphology
in limbward viewing is a distinctive \acp{EB} characteristic.
It was used in \PaperII\ to establish \acp{EB} brightness criteria,
but these may fail close to disk center (\PaperIII) where
differentiation with the \acp{MC}s that produce \acp{EB}s (as in the
serpentine U-loop pull-up scenario of \acp{EB} formation of \eg\
\citeads{2002SoPh..209..119B}, 
\citeads{2004ApJ...614.1099P}, 
\citeads{2007ApJ...657L..53I}, 
\citeads{2009A&A...508.1469A}, 
\citeads{2009ApJ...701.1911P}) 
is less easy, as already pointed out by
\citetads{1917ApJ....46..298E} 
and later by \citetads{1977ASSL...69.....B}. 

With ``FAF'' I denote small sudden brightenings in \acp{AIA}
1600\,\AA\ images with filamentary morphology.  
Similarly to \acp{EB}s they appear as sudden
brightenings in complex active regions, but they are elongated,
change faster, and show rapid apparent motion.
They seem to start as \acp{EB}-like reconnection events but then break
through the chromospheric canopy and affect the higher atmosphere
(\PaperIII).
\citetads{2009ApJ...701.1911P} 
noted them in 1600\,\AA\ images from \acp{TRACE}, called them
``transient loops'', and reported them as a new phenomenon -- but
probably \Halpha-core ``microflares'' (\eg\
\citeads{2002ApJ...574.1074S}) 
describe similar outbursts.
In \citetads{2013JPhCS.440a2007R} 
we noted them as ``small flaring arch filaments and microflares'' and
abbreviated this to FAF = ``flaring arch filament'' in \PaperIII, but
a better name is ``flaring active-region fibril'' to avoid confusion
with the larger and stabler structures making up ``arch filament
systems'' in emerging active regions.

\acp{IRIS} bombs (IB) are short-lived small-scale active-region
brightenings that show very bright \acp{IRIS} \CII\ and \SiIV\ lines
with very broad doubly peaked profiles and superimposed absorption
blends indicating cool overlying gas.
In their discovery paper \citetads{2014Sci...346C.315P} suggested that
these are pockets of hot gas in the upper photosphere and may
correspond to \acp{EB}s in \Halpha.
In \PaperIII\ we found that they are more likely \acp{FAF}s while
\acp{EB}s may also show \acp{IB} signatures in their aftermaths. 

\paragraph{Photosphere, chromosphere, clapotisphere.}
%%%%%%%%%%%%%%%%%%%%%%%%%%%%%%%%%%%%%%%%%%%%%%%%
\PaperI\ concluded that \acp{EB}s are a photospheric phenomenon,
but also that they reach up to one Mm or more.  
A one-Mm tall feature embedded in the ALC7 atmosphere would have its
top in the ALC7 chromosphere; that part would then be called
chromospheric.  
However, we designated \acp{EB}s instead as fully non-chromospheric,
hence implicitly photospheric, because even their tops always
remain shielded by the fibril canopy observed in \Halpha\ line center
in active regions. 
What we meant is that \acp{EB}s are embedded in cool gas with
upper-photosphere temperatures, as present between shocks in the
internetwork areas of the HION atmosphere in Fig.~\ref{fig:HION} where
it reaches 3\,Mm height, higher than \acp{EB}s.

However, recently Ph.~Judge (private communication) insisted
that one should restrict ``photosphere'' to its original meaning: the
domain where the bulk of the solar radiation flux escapes.
I follow his admonishment here and restrict ``photosphere'' to
this thin shell which reaches no higher than about 400\,km above
$\tau_{5000}\is 1$.
Its upper layers are similar in the ALC7, HION and FCHHT-B
atmospheres, with a temperature decline that is also closely the same
in \acp{LTE} radiative-equilibrium models.
The reason is that this is the most homogeneous part of the solar
internetwork atmosphere, above the granulation and below shock
formation and not too magnetized; linear undulations of the acoustic
$p$-mode pattern and internal gravity waves represent its major
perturbations.
The visible continuum escape is described fairly well by \acp{LTE} as
a small leak from a large thermal pool.
Theoretical radiative-equilibrium modeling 
and empirical 1D static modeling 
apply best to the upper photosphere.

%% fig:CE-LTE
%===========================================================================
\begin{figure}
  \centerline{%
    \includegraphics[width=\columnwidth]{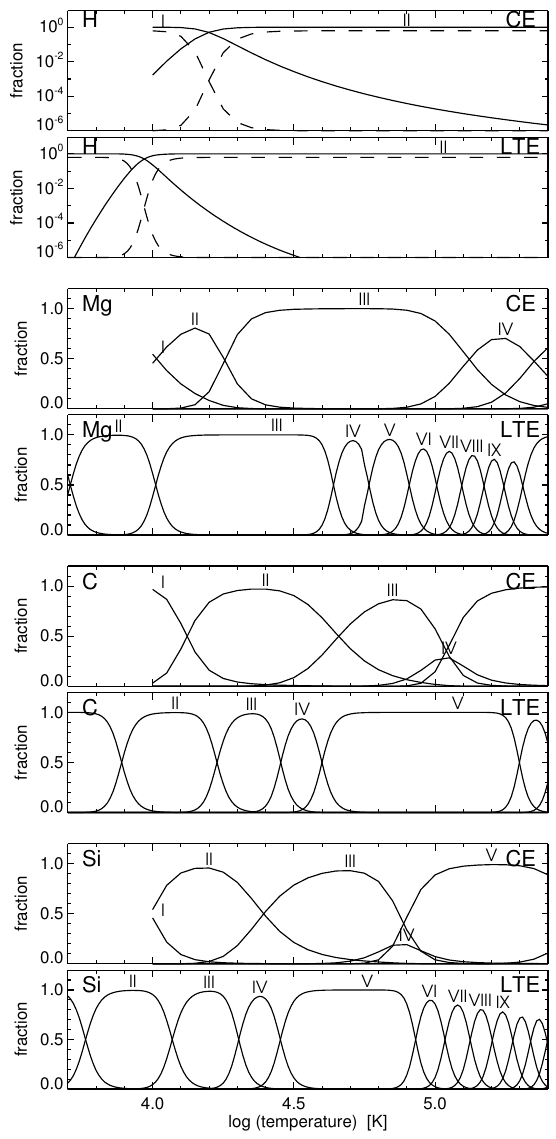}}
    \caption[]{\label{fig:CE-LTE} 
    \acp{CE}/\acp{LTE} ionization-stage population comparisons for H,
    Mg, C, and Si.
    {\em Upper panel of each pair\/}: \acp{CE} distribution with
    temperature (CHIANTI gives no values below 10\,000\,K).
    {\em Lower panel of each pair\/}: \acp{LTE} distribution with
    temperature for fixed electron density
    $N_\rme \is 10^{14}$~cm$^{-3}$.
    For 10 times lower $N_\rme$ the \acp{LTE} curve patterns remain
    similar but the peaks shift leftward over about $-0.05$ in $\log(T)$
    and become narrower.
    The first pair for hydrogen has logarithmic $y$-axes; the dashed
    curves are on the linear scales of the other panels.  
    {\em Remarkable\/}: the substantial leftward \acp{CE} to \acp{LTE}
    shifts and the wide presences of \MgIII, \CV, and \SiV.
    }
\end{figure}
%===========================================================================

The name ``chromosphere'' was given by
\citetads{1868RSPS...17..131L} 
to the pink shell he saw spectroscopically (literally) around the sun,
with the beautiful color due to emission in \Halpha, \Hbeta, and
\HeIDthree.
It then became the name of what causes the flash spectrum
(\citeads{1961psc..book.....A}), 
but in the past decades it became synonymous with the raised
temperature plateau over $h \approx 500-2000$\,km in
Avrett-style standard models as ALC7.

I prefer to return to Lockyer's naming and therefore define as on-disk
chromosphere what one observes in \Halpha. 
In active regions this is a dense canopy of long opaque fibrils.
I sketched one as example in Fig.~\ref{fig:HION} because HION does not
contain such fibrils (an issue addressed in the next paper in this
series).
Note that the green arches in Fig.~\ref{fig:HION}, which resulted from
photon suction by \Halpha\ where hydrogen ionizes, are artifacts
because the HION computation did not include photon losses in the
Lyman lines.
In HION the \HI\ $n\is2$ level therefore acted effectively as ground
state for the \HI\ atom top. 

With these definitions a third name is needed for shocked cool gas in
high-reaching sub-canopy domains in solar internetwork areas and also
present in HION internetwork (Fig.~\ref{fig:HION}).
I once again use ``clapotisphere''
(\citeads{1995ESASP.376a.151R}). 

Thus, a major difference between the ALC7 and HION atmospheres is that
ALC7 possesses a static 1D isothermal chromosphere, HION a highly
dynamic low-lying 2D network chromosphere made up of dynamic fibrils
and a highly dynamic high-reaching 2D internetwork clapotisphere.
This difference is enormous.  

In my opinion HION comes much closer to the actual Sun even while it
is only 2D and has no internetwork-covering \Halpha\ fibrils.
I regard the temperature stratification of ALC7-like ``standard''
chromospheres, for all their phenomenal didactic value, 
as artifacts from unrealistic static modeling.
They do not describe a mean over actual temperature fluctuations
because their construction through ultraviolet spectrum fitting
suffers from non-linear Wien temperature sensitivity and therefore
favors shocks (Fig.~4 of
\citeads{1994chdy.conf...47C}.) 
Indeed, the ALC7 chromosphere has similar temperature, ionization and
electron density as the HION shocks.

Worse, while every solar-atmosphere column must have a minimum
temperature and a steep rise to the corona somewhere, the HION
atmosphere suggests that their heights fluctuate so wildly that the
static-modeling notions of ``the temperature minimum'' and ``the
transition region'' are useless.

%%%%%%%%%%%%%%%%%%%%%%%%%%%%%%%%%%%%%%%%%%%%%%%%%%%%%%%%%%%%%%%%%%%%%%%%%%%%
\section{Analysis}    \label{sec:analysis}
%%%%%%%%%%%%%%%%%%%%%%%%%%%%%%%%%%%%%%%%%%%%%%%%%%%%%%%%%%%%%%%%%%%%%%%%%%%%

\subsection{LTE--CE comparisons} \label{sec:LTE-CE}
%%%%%%%%%%%%%%%%%%%%%%%%%%%%%%%%%%
Figure~\ref{fig:CE-LTE} compares the fractional populations of
ionization stages of H, Mg, C and Si between \acp{CE} (upper panels per
pair) and \acp{LTE} (lower panels).

The upper panels are from CHIANTI and represent ``Carole Jordan''
diagrams following
\citetads{1969MNRAS.142..501J} 
whose landmark Fig.~3 represents the solar-corona counterpart to
Payne's stellar-photosphere population distributions.

The lower panels of Fig.~\ref{fig:CE-LTE} are not strictly ``Cecilia
Payne'' diagrams because they do not include Boltzmann excitation for
particular lines but show only the distribution over successive
ionization stages defined by the Saha equation.
They are also based on CHIANTI data (ionization energies and term
structures defining partition functions).
The large electron density corresponds to full hydrogen
ionization of ALC7 gas at $h\is850$\,km.

These extreme assumptions bracket the temperature regime in which the
\acp{EB} diagnostics form. 
Of course neither is likely valid, but here I suggest that \acp{EB}
onset visibilities, even in the \acp{IRIS} lines, behave more as the
lower \acp{LTE} panel than the upper \acp{CE} panel per pair.

The peak widths are defined by the ionization energies and are
exceptionally large for the closed-shell ions \MgIII, \CV, and \SiV,
and of course for \HII\ which suffers no competition from higher stages.

The hydrogen curves (first pair of panels) are plotted on logarithmic
scales in view of the enormous hydrogen abundance: even at only
$N_{\tiny \HI}/N_{\rm total} \is 10^{-5}$ neutral hydrogen still
competes with the dominant stages of the most abundant metals. 
For \acp{CE} such presence extends as high as 100\,000\,K.
For \acp{LTE} the break-even ionization point shifts below 10\,000\,K.

The next pair (Mg) shows that due to the similarity of the \MgII\ and
\HI\ ionization energies (15.0 and 13.6 eV, respectively) the \MgII\
to \MgIII\ transition occurs very similar to \HI\ to \HII\ ionization.
Hence, the resonance \MgII\ \hk\ and \HI\ Lyman lines will have
similar presence apart from the 2.5\,10$^4$ extinction ratio per given
feature (abundance ratio), unless they also differ in being closer to
the \acp{LTE} limit or to the \acp{CE} limit or in departure from
\acp{SE}.

The third and fourth pairs are for C and Si. 
They show similar patterns, with wide peaks for the first and
especially the fourth ions.
In \acp{CE} the \SiIV\ presence peaks with relatively low amplitude at
$\log(T)\!\approx\!4.9$ ($T\!\approx\!80000$~K) as quoted by
\citetads{2014Sci...346C.315P}, 
but in \acp{LTE} it peaks already at $T\!\approx\!25\,000$~K and also
with much larger amplitude.
Since the \SiIV\ lines in the \acp{EB} spectra of \PaperIII\ reach
only \acp{EB} thickness $\tau\!\approx\!1$ they may be formed on
the rising branch and represent only $T\!\approx\!15\,000-20\,000$~K.

%% fig:extLTE
%===========================================================================
\begin{figure*}
  \sidecaption
  \includegraphics[width=120mm]{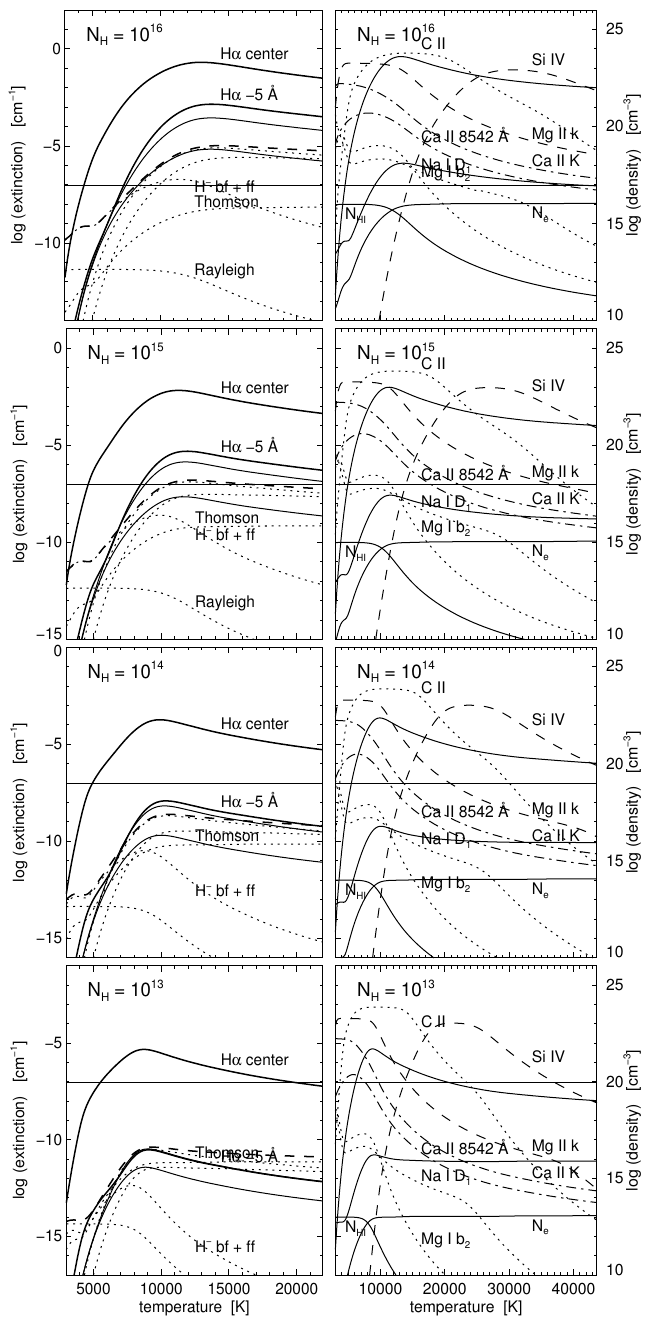}
  \caption[]{ \label{fig:extLTE} 
  Payne-style extinction comparisons for solar gas assuming \acp{LTE}. \\
  {\em Top to bottom\/}: total hydrogen density
  $N_{\rmH}\is10^{16}, 10^{15}, 10^{14}, 10^{13}$~cm$^{-3}$,
  corresponding to heights 350, 615, 850 and 1150\,km in ALC7
  (Fig.~\ref{fig:ALC7}).
  The logarithmic extinction scales along the $y$-axes shift
  upward between panels to compensate for the density decrease.
  The horizontal line at $y\is-7$ marks optical thickness unity for a
  slab of 100~km geometrical thickness.\\
  {\em Left\/}: the upper two solid curves show \Halpha\ extinction at
  line center and in the wing at $\Delta \lambda \is -5$~\AA.  
  The third solid curve is for $\Delta \lambda \is -5$~\AA\ when using
  a Voigt instead of a Holtsmark distribution for linear Stark
  broadening.
  The lowest solid curve results when Stark broadening is neglected.
  The dashed curve is the total continuous extinction at the \Halpha\
  wavelength, made up by \Hmin\ and \HI\ bound-free and free-free
  transitions and by Thomson and Rayleigh scattering.
  The dotted curves show the separate continuum contributions (the top
  two for \HI\ bound-free and free-free extinction are not labeled).\\
  {\em Right\/}: the \Halpha\ line-center extinction is shown again
  (uppermost solid curve) but over a doubled temperature range.
  It is compared with the line-center extinction of \MgIIk\ (dashed),
  \CaIIK\ and \CaIR\ (dot-dashed), \MgIbtwo\ and \NaIDone\ (dotted),
  and the \acp{IRIS} lines \CII\ 1335.71\,\AA\ (dotted) and \SiIV\
  1393.75\,\AA\ (dashed).
  The second solid curve is the total continuous extinction at
  1700\,\AA\ due to \Hmin, \HI\ and continuous scattering.
  The two solid curves underneath show the competing neutral hydrogen
  density $N_{\tiny \HI}$ and free electron density $N_\rme$ on a
  comparable (but non-shifting) logarithmic scale specified along the
  $y$-axes at right.\\
  {\it Remarkable\/}: at left the large effect of Stark broadening.
  At right the large \Halpha\ extinction at high temperature.
  \vspace{2ex}\mbox{}} 
\end{figure*}
%===========================================================================

\subsection{\Halpha\ extinction in LTE} \label{sec:Ha-LTE}
%%%%%%%%%%%%%%%%%%%%%%%%%%%%%%%%%%%%%%
Figure~\ref{fig:extLTE} shows extinction-versus-temperature diagrams
that I wish to call ``Cecilia Payne diagrams'' although they are
somewhat more elaborate than Payne's curves. 
First, they show the \acp{LTE} line extinction coefficient rather than
the lower-level Saha-Boltzmann population by multiplying with the
transition probability, the line profile, and the correction for
stimulated emission (with $b_l\is b_u\is 1$ in Eq.~\ref{eq:ext}).
Second, they do not show curves for fixed electron pressure but
for fixed total hydrogen density $N_\rmH$ (a parcel of
solar gas in which the electron density varies with temperature
according to the degree of ionization, especially of hydrogen).

$N_\rmH$ diminishes in ten-fold steps from top to bottom.
In the ALC7 atmosphere these values correspond to heights
350, 615, 850, and 1150\,km, respectively, sampling its upper
photosphere and lower chromosphere (Fig.~\ref{fig:ALC7}).
The extinction scales along the lefthand $y$-axes shift accordingly.

The horizontal line at $y=-7$ corresponds to optical thickness unity
along 100~km path length and schematically represents an \acp{EB} in
sideways viewing. 
Above this line a 100~km slab appears optically thick, below it
optically thin.

This section discusses the first column.
It shows comparisons between \Halpha\ and the continuum.
The various curves were computed using the formulae and tables in
\citetads{2005oasp.book.....G} 
for the element mix of
\citetads{2009ARA&A..47..481A}. 
For \Halpha\ linear Stark broadening was included using Gray's
equation (11.49) with the first approximation of
\citetads{1978JQSRT..20..333S} 
for the Holtsmark distribution.
The Saha equation was iterated for hydrogen, helium and the first two
ionization stages of the abundant metals in order to obtain the
free-electron density (shown at right).

The comparisons in the first column of Fig.~\ref{fig:extLTE} were
prompted by the suggestion of
\citetads{1968mmsf.conf..109E} 
that the frequency redistribution which occurs in thermal electron
scattering may cause the extraordinary \acp{EB} moustaches by shifting
\Halpha\ line-core photons over large electron Dopplershift into the
far \Halpha\ wings.
They stipulated that for this mechanism \acp{EB}s must contain sufficient
free electrons to offset the very small Thomson cross-section
$\sigma_{\rm T} = 6.65\,10^{-25}$\,cm$^2$, quantifying that electron
density $N_\rme \is 10^{16}$~cm$^{-3}$ would be needed along a 1000-km
line of sight through an \acp{EB} to gain optical thickness of order
unity.
They suggested that this may be reached through hydrogen ionization in
photospheric gas.

My initial enthusiasm for this mechanism (Sect.~6 of
\citeads{2013JPhCS.440a2007R}) 
was undone by the tests shown here, based on similar tests by
J.~Leenaarts (private communication).
Neutral hydrogen extinction in the Paschen continuum takes over from
\Hmin\ extinction already below $T\is10\,000$~K, whereas Thomson
scattering becomes the dominant continuum agent above 10\,000\,K only
below hydrogen density $N_\rmH \approx 10^{13}$\,cm$^{-3}$ (bottom
panel) with extinction below $10^{-10}$\,cm$^{-1}$ so that only one
per thousand \Halpha\ photons would be Thomson-scattered and
Doppler-shifted in a 100-km \acp{EB} slab. 
For more scattering the density must be higher, but then such a slab
becomes opaque in the Paschen continuum.

In the ALC7 chromosphere the Paschen extinction drops significantly below
the \acp{LTE} value (dashed curve in the first panel of
Fig.~\ref{fig:ALC7}) due to photon losses in \Halpha, but in HION this
happens only in the green arches in Fig.~\ref{fig:HION}, not in the
shocks or the post-shock cool clouds.
Therefore, the graphs in Fig.~\ref{fig:extLTE} suggest that \acp{EB}s
would be observed in the continuum well before Thomson scattering
would cause moustaches, making the suggestion of
\citetads{1968mmsf.conf..109E} 
incompatible with the observed \acp{EB} transparency in the continuum.

In addition, other lines such as \CaIIH\ and \CaIR\ do show \acp{EB}
moustaches but less wide than \Halpha\ 
(\citeads{2010PASJ...62..879H}; 
private communication from R.~Rezaei).
For Thomson redistribution they should have similar extent since it
acts the same for any line with similar core brightening below the
fibril canopy.

Fortunately, the first column of Fig.~\ref{fig:extLTE} also suggests a
more viable \acp{EB} moustache mechanism: linear Stark broadening by
electrons.
In the second panel a 100-km slab is optically thick (extinction above
10$^{-7}$\,cm$^{-1}$) in the \Halpha\ wing at
$\Delta \lambda \is -5$~\AA\ from line center at all temperatures
above 8000\,K, while it becomes only barely opaque in the continuum at
$T \is 11\,000-15\,000$~K.  
This parameter domain may therefore explain \acp{EB} visibility in the
\Halpha\ wing together with transparency in the continuum.
However, it is essential that linear Stark broadening, a specific
agent for Balmer lines,
is included in the line formation computation since without it the
\Halpha\ wing has less extinction than the continuum.
This is demonstrated by the lowest solid curve, which results from
only accounting for radiation damping and Van der Waals broadening as
done by \citetads{2013A&A...557A.102B}. 

The third solid curve from the top in each lefthand panel results when
a Voigt function is used for the linear Stark broadening (as is done
in the RH code) instead of the Holtsmark distribution.
It lies slightly lower due to the slower Lorentzian
$\sim\!(\Delta \lambda)^{-2}$ wing decay then Holtsmark
$\sim\!(\Delta \lambda)^{-5/2}$ decay, which through area
normalization produces profile cross-overs in the wings further out
than $\Delta \lambda \is -5$~\AA.

The sampling wavelength $\Delta \lambda \is -5$~\AA\ of these
comparisons was selected because the observed \acp{EB} excess spectrum in
Fig.~2 of \citetads{1983SoPh...87..135K} 
suggests the start of optical thinness at about this wavelength for
slab-like profile formation.
In this \acp{EB} the moustache extent was over 10~\AA\ on each side of line
center.
Note that \citetads{1917ApJ....46..298E} 
specified bomb extents up to $\Delta \lambda \is 15$~\AA; so did
\citetads{1956Obs....76..241S} 
for moustaches.

Thus, linear Stark broadening seems a viable mechanism to transfer
\Halpha\ photons from the locally bright but obscured line core to
escape in the outer wings and pass through overlying fibrils.
Similarly to Thomson redistribution, it requires sufficient hydrogen
ionization but it has much larger cross-section.
I identify it therefore as a key agent producing bright extended \Halpha\
moustaches and making the Balmer lines display the \acp{EB} phenomenon so
spectacularly.

\subsection{LTE extinction of EB diagnostics} \label{sec:others-LTE}
%%%%%%%%%%%%%%%%%%%%%%%%%%%%%%%%%%%%%%%%%%%%
I now turn to the second column of Fig.~\ref{fig:extLTE}.
It compares the \acp{LTE} extinction of \Halpha\ with that of other
\acp{EB} diagnostics, again as function of temperature but over a
larger temperature range.
The different curves show \acp{LTE} line-center extinction for the
specified lines and the Balmer continuum extinction at 1700\,\AA,
defined by their Boltzmann and Saha sensitivities to temperature and
electron density.
The electron and neutral hydrogen densities are overplotted in the
bottom part of each panel, with axis scales at right.

The little plateau at the start of the $N_\rme$ curve in the top
panel, also evident in the 1700\,\AA\ curve, illustrates that
ionization of abundant metal atoms provides the electrons for \Hmin\
extinction in the low-temperature photosphere. 
At higher temperature the electrons come mostly from hydrogen.
The cross-over point with 50\% hydrogen ionization and
$N_{\tiny \HI}\!\approx\!N_\rme$ shifts from $T\!\approx\!12\,000$\,K
to $T\!\approx\!8\,000$\,K from top to bottom.
The curves and their cross-over temperature values change
correspondingly between panels, but not drastically.

In the lefthand column the second panel suggested
$\log(N_\rmH)\!\approx\!15$ and $T\!\ga\!10\,000$\,K as viable target
parameters to obtain \Halpha\ wing visibility combined with continuum
transparency for the 100-km slab.   
The righthand column permits such \acp{LTE} visibility estimation also
for the other \acp{EB} diagnostics.
I discuss them one by one.

\paragraph{\Halpha.}
%%%%%%%%%%%%%%%%%%
The uppermost solid curve in each panel is again the \Halpha\
line-center extinction, identical to the corresponding curves at left.  
The most striking feature in these graphs is the discordant behavior
of \Halpha\ due to its combination of exceedingly large elemental
abundance and exceedingly large excitation energy.
In each graph the \Halpha\ line-center extinction shows a steep rise
with temperature due to its Boltzmann population factor, followed by a
decline due to hydrogen ionization which is considerably tempered by
the ever-increasing relative excitation.
At low temperature \Halpha\ has no significant extinction whatsoever,
but above $T \!\approx\! 8000$~K \Halpha\ becomes much stronger than
\CaIIK, the reverse of their photospheric ratio, and \Halpha\ even
beats \MgIIk\ significantly above $T \is 12\,000$~K.
This extraordinary strength may actually yet increase through the much
slower decay for \acp{CE} ionization (Fig.~\ref{fig:CE-LTE}) combined
with Boltzmann excitation forcing by \Lyalpha. 

A numerical fit of the locations of the peaks of the \Halpha\
extinction curves, which are defined by the sensitivities of the
Boltzmann, Saha, profile shape and induced emission terms in
Eq.~\ref{eq:ext}, shows that these obey:
\begin{equation}
    \log (n_2/N_\rmH) = 0.683\,\log(N_\rmH)-14.8,
    \label{eq:peakfit}
\end{equation}
which provides a quick estimate of the maximum extinction that 
\Halpha\ can reach in dynamical situations with onset \acp{LTE}
validity.

The \Halpha\ and \CaIR\ curves cross at
$T \!\approx\! 7\,000-8\,000$\,K. 
The 100-km slab is optically thick at this curve cross-over in
all panels, conform the invisibility of \acp{EB}s at the centers of
these lines from obscuration by the fibrilar canopy (\PaperI).

\Halpha's special combination of very large high-temperature
extinction with very small low-temperature extinction also
implies that a high-temperature slab embedded in cool gas can display
large \Halpha\ wing emissivity within a transparent environment,
without shielding along lines of sight as C and B in
Fig.~\ref{fig:HION}.
Fig.~\ref{fig:extLTE} suggests that cool surrounding gas will not have
any \Halpha\ opacity. 
The fractional population curve (dashed) in the first panel of
Fig.~\ref{fig:ALC7} shows that in the ALC7 atmosphere the \Halpha\
opacity is indeed negligible between the deep photosphere
and the chromospheric plateau.

Close to an \acp{EB} absorption of \Lyalpha\ photons from the \acp{EB}
itself boosts the surrounding $n\is2$ population as demonstrated in
Sect.~\ref{sec:spatial}, but the extent of such halos is small and
they occur only in the \Halpha\ core in which the halo is invisible
anyhow from obscuration by overlying fibrils.
In the outer \Halpha\ wings surrounding cool gas lacks moustache
opacity from shortage of free electrons for Stark broadening.  
The therefore unmolested \Halpha-moustache sharpness permits, at
sufficiently high angular resolution, to resolve very fine
substructure within \acp{EB}s, as observed in
\citetads{2010PASJ...62..879H} and \PaperI.

Also, the hotter \acp{EB} top observed along line of sight A retains
large \Halpha\ emissivity thanks to the slow extinction decline
for increasing temperature. 

\paragraph{Balmer continuum.}
%%%%%%%%%%%%%%%%%%%%%%%%%%%%
The second solid curve is the continuous extinction due to hydrogen
and continuous (Thomson and Rayleigh) scattering at 1700\,\AA. 
In the top panels of Fig.~\ref{fig:extLTE} these curves mimic
the \Halpha\ curve because both are set by the hydrogen $n\is2$
population.
In the lower panels Thomson extinction flattens the 1700\,\AA\ curves,
similarly to the dashed 6563\,\AA\ continuum curves (with Paschen
instead of Balmer extinction) in the lefthand column.
The 1700\,\AA\ curve peaks just below the $\tau\is1$ line for the
100-km slab in the second panel; thickness requires a wider slab or
larger density.  

Below 10\,000\,K the 1700\,\AA\ extinction is dominated by bound-free
\SiI\ and \FeI\ edges, but at higher temperature the main agent is the
Balmer continuum.
This is already evident in the classic \acp{LTE} continuum extinction
diagrams (in German) in Figs.~I--XV of
\citetads{1951ZA.....28...81V}, 
the third landmark thesis by a heroine of astrophysics emulated here.
They have been redrawn (in English and with better graphics) in
Figs.~3-12A--O of \citetads{1973itsa.book.....N}. 
I do not include the cool-gas \SiI\ and \FeI\ contributions here
because \acp{LTE} would give severe overestimation also for
surrounding gas.
They are diminished through photoionization already in the upper
photosphere of 1D standard models
(\citeads{1981ApJS...45..635V}) 
and are further depleted near hot \acp{EB}s through irradiation by
these. 

The ultraviolet continua are all strongly resonant-scattering, as
evident from the superb formation diagrams in Fig.~36 of
\citetads{1981ApJS...45..635V}. 
In their VALIIIC model the Eddington-Barbier $\tau\is1$ heights around
1700\,\AA\ range over 300--500\,km, but most photons are thermally
created in the deep photosphere and then scatter out with a
$S\!\approx\!J$ decline which, opposite to the line scattering in
Fig.~\ref{fig:ALC7}, has $J > B$ from the combination of temperature
gradient setting by optical photon losses, deep formation, large Wien
temperature sensitivity, and the gradient sensitivity of the Lambda
operator (Chapt.~4 of \RTSA).  
A deeply located temperature increase boosts $J$ and the emergent
continuum intensity across the ultraviolet while an increase near
$\tau\is1$ does not.
Even along line of sight C in Fig.~\ref{fig:HION} the bright foot of
an \acp{EB} so has ultraviolet visibility, be it diffuse from
surround scattering by the remaining Si and Fe atoms and in the
Balmer continuum boosted by \Lyalpha\ irradiation.

Large depth provides the density needed for noticeable
Balmer-continuum brightening (Fig.~\ref{fig:extLTE}).
In addition, since \acp{EB}s are bi-modal jets, compression at the
lower shock likely enhances the local Balmer-continuum production.

The 1700\,\AA--\Halpha-wing comparisons in Fig.~6 of \PaperII,
Fig.~4 of \citetads{2013JPhCS.440a2007R}, 
and Figs.~2, 6, 9 of \PaperIII\ indeed suggest, even at the low
\acp{AIA} resolution, that \acp{EB}s appear more extended in the
1700\,\AA\ images than in the \Halpha-wing images, that the
1700\,\AA\ images favor \acp{EB} feet, and that these show the
largest blurring.
The \acp{IRIS} spectra in \PaperIV\ indeed show \acp{EB}
brightening in the 1400\,\AA\ continuum. 

\paragraph{\CaII\ lines.}
%%%%%%%%%%%%%%%%%%%%
\CaIIK\ and \CaIR\ have similar curves in Fig.~\ref{fig:extLTE}
except for the Boltzmann sensitivity of the latter. 
In both lines the slab is thick for the target parameters, but
overlying cooler fibrils are yet thicker so that only wing
brightenings result, at smaller moustache extent then for \Halpha\
since these lines suffer no Stark broadening.
This is also conform observations (\PaperII\ for \CaIR,
\citeads{2010PASJ...62..879H} 
for \CaIIH).

At temperatures above their cross-overs in Fig.~\ref{fig:extLTE}
\Halpha\ and \CaIR\ have opposite temperature sensitivity, implying
that the \CaIR\ wings favor cooler \acp{EB} feet, the
\Halpha\ wings hotter \acp{EB} tops. 
This is seen in Fig.~3 of \PaperII. 
In addition, the downward-directed part of a bi-modal \acp{EB} jet
favors blue-wing emission, also evident in that figure.
Furthermore, the larger \CaII\ extinction at lower temperature implies
scattering in surrounding cooler gas causing defocus, also evident in
that figure.
For \CaIIH\ diffuse \acp{EB} halos were reported by
\citetads{2008PASJ...60..577M}. 

\paragraph{\NaIDone\ and \MgIbtwo.} 
%%%%%%%%%%%%%%%%%%%%%%%%%%%%%%%%
These lines have cross-over extinction curves in Fig.~\ref{fig:extLTE}
(dotted), with \MgIbtwo\ stronger below
$T\!\approx\!10\,000$\,K. Both curves decay above
10\,000\,K from increasing ionization, but less steep for
\NaIDone\ due to the high \NaII\ ionization energy (47.3~eV,
highest after Li and He.
The initial steep declines are also from increasing ionization, but
over 4\,000--10\,000\,K ionization diminishes due to the
increasing electron density.

In the $N_\rmH \is 10^{15}$ panel the 100-km slab reaches about
$\tau\!\approx\!1$ in both line centers near $T\!\approx\!10\,000$\,K,
but remains less opaque than in gas with $T\!\approx\!4000$\,K or
less around \acp{EB}s.
Photon suction as in ALC7 enhances cool-gas
extinction beyond the \acp{LTE} estimate.
Obscuration along lines of sight as C and B in
Fig.~\ref{fig:HION} is therefore probable.
\acp{EB}s are indeed not observed in the cores of these lines, in
exceptional cases only in the wings of \MgIbtwo\ (\PaperIV).
For this line ultraviolet \MgI\ overionization may offset the suction
increase, as in Fig.~\ref{fig:ALC7} (compare the $b_l$ curves of the
two lines).

\paragraph{IRIS lines.}
%%%%%%%%%%%%%%%%%%%%
For the ultraviolet \acp{IRIS} lines the second panel predicts
enormous slab opacity in the \CII\ and \MgIIk\ lines for
$T \approx 10\,000$\,K, but invisibility for the \SiIV\ lines which
actually showed \acp{EB}s ``thinnish'' ($\tau\!\approx\!1$) in
\PaperIII.    
That value is reached at $T\!\approx\!15\,000$\,K, towards the
\SiIV\ curve peaks around $T\is25\,000$\,K as in
Fig.~\ref{fig:CE-LTE}.
It does so in all panels since the leftward shifts of the \SiIV\ curve
for lower electron density are offset by the downward shifts for lower
gas density (upward $y$-axis shifts) in sampling the steep curve
increases.
The assumption of \acp{LTE} ionization is least likely for this
diagnostic, but departures shift the curves to higher temperature as
in Fig.~\ref{fig:CE-LTE}; $T\!\approx\!15\,000$\,K represents a
minimum estimate.

One might expect that such high temperatures are only reached
in \acp{EB} tops, but the \acp{IRIS} spectra in \PaperIII\ also show
enhanced-intensity redshifted \SiIV\ profiles with the slitjaw images
confirming that these come from lower \acp{EB} parts.
The blueshifted profiles from the upper parts are brighter and
wider, but nevertheless also \acp{EB} feet show up.

Surrounding gas can also affect the ultraviolet \acp{EB} signatures. 
Most spectra in \PaperIII\ contain absorption blends:
\MnI\ blends on \MgIIhk\ without Dopplershifts,
\FeII\ and \TiII\ blends on the \SiIV\ and \CII\ lines with
10\,\kms\ blueshifts. 
The first were attributed to undisturbed upper-photosphere gas along
lines of sight as C in Fig.~\ref{fig:HION}, the latter to
upper-clapotisphere gas containing shocks along lines of sight as B.
Lines of sight as A then sample hot \acp{EB} tops without such blends.

These blends and their small Dopplershifts, observed not only in
\acp{EB}s but also in \acp{FAF}s (\PaperIII), support the
conclusion of \citetads{2014Sci...346C.315P} 
that \acp{IB}s occur at large depths.  
For \acp{EB}s this was already established in \PaperI: they have their
feet in the low photosphere.
Comparison of the IRIS spectra and cutout images in \PaperIII\ shows
that profiles with deep \MnI\ blends correspond to lower-part
samplings with the largest 1700\,\AA\ (\acp{EB}s) or
1600\,\AA\ (\acp{FAF}s) halos, suggesting photospheric formation even
within the narrow Judge sense adopted in Sect.~\ref{sec:terminology}.

\acp{EB}s appear as sharp in the \SiIV\ lines as in \Halpha\
moustaches: without obscuration by overlying fibrils and without
defocusing by surrounding gas.
The nominal continuum at 1400\,\AA\ has Balmer brightening in
\acp{EB}s with scattering halos comparable to 1700\,\AA, but the
\SiIV\ lines from \acp{EB}s are much brighter and dominate the
\acp{IRIS} 1400\,\AA\ slitjaw passband (\PaperIII).
The lines from a deep \acp{EB} foot suffer some surround scattering as
observed in the blurred 1700\,\AA\ signature, either in the \SiI\
edge that contributes most continuous opacity at 1400\,\AA\ in cool
gas but is depleted by irradiation from the \acp{EB} or in the Balmer
edge boosted by \Lyalpha\ irradiation.
However, even if a sizable fraction of the \SiIV\ emission from an
\acp{EB} foot suffers such surround scattering, this undergoes complete
redistribution over the edge with most re-emission near its threshold.
It causes only gray attenuation of the lines, not affecting their
relative profile shape or image sharpness.
There is indeed good size and morphology likeness between the
1400\,\AA\ and \Halpha-wing images in Figs.~2, 6 and 9
of \PaperIII.

\subsection{EB density and temperature estimation} \label{sec:locus}
%%%%%%%%%%%%%%%%%%%%%%%%%%%%%%%%%%%%%%%%%%%%%%%%%%%%
This section discusses how the graphs in Fig.~\ref{fig:extLTE}
constrain the density and temperature parameter space of \acp{EB}
onsets assuming the \acp{LTE} recipes of Sect.~\ref{sec:background}.

The second column of Fig.~\ref{fig:extLTE} shows only line-center
extinctions to minimize figure clutter.
However, as for \Halpha\ in the lefthand column, not the line-center
but the wing extinction should be estimated for those diagnostics that
only show \acp{EB}s in their wings -- all of them except the \SiIV\
lines and the 1700\,\AA\ continuum. 
All curves except these two should therefore be shifted downward over
1--2 log units. 
At 10\,000\,K \NaIDone\ and \MgIbtwo\ then become weaker than the
1700\,\AA\ continuum, in agreement with the actual \acp{EB}
visibilities.

However, these are not the only shifts to be applied.
An obvious one is to shift the $y\is-7$ line vertically by selecting
different slab thickness. 
My arbitrary choice of 100\,km was inspired by typical diameters of
network \acp{MC}s.
A thinner slab requires higher density to reach $\tau\is1$.
%%and density change from panel to panel has the opposite effect.

Another shifting agent is extinction memory of hot prior moments, as
the Balmer lines and the Balmer continuum possess in HION shock
aftermaths. 
If hot and dense \acp{EB} onsets are also followed by cooler, more
tenuous aftermaths then similar delays in population adjustment may
occur.
Such memory then implies leftward shifts of the \Halpha\ and
1700\,\AA\ extinction curves, retaining the extinction of the hotter
moment in the past into the cooler present.
Similar memory shifts may also occur for the \SiIV\ and \CII\ lines.

An example scenario is that \acp{EB} tops heat to 20\,000\,K in
their onset, gain $\tau\!\approx\!1$ opacity (and
corresponding emissivity) in the \SiIV\ lines, and retain this
as well as high Balmer line-wing and continuum emissivity during
subsequent cooling.

Furthermore, \acp{EB}s are not homogeneous slabs but appear hotter
higher up (\PaperIII).
The curve patterns in Fig.~\ref{fig:extLTE} illustrate that the
optical lines except \Halpha\ favor cooler and denser gas than the
\acp{IRIS} lines.
The former are more likely to sample the lower parts of an \acp{EB} as
along lines of sight C to B in Fig.~\ref{fig:HION}, the latter the
hotter and more tenuous upper parts along lines of sight B to A.
\Halpha\ is special in showing the full \acp{EB} extent.

The upshot is that Fig.~\ref{fig:extLTE} does not define a single
[temperature, density] locus to pinpoint all \acp{EB} visibilities, but
with these various curve-shifting options it does suggest the domain
where they originate: hydrogen density around 10$^{15}$\,cm$^{-3}$,
time-dependent temperatures in the 10\,000-20\,000\,K range. 

%%%%%%%%%%%%%%%%%%%%%%%%%%%%%%%%%%%%%%%%%%%%%%%%%%%%%%%%%%%%%%%%%%%%%%%%%%%%
\section{Discussion}\label{sec:discussion}
%%%%%%%%%%%%%%%%%%%%%%%%%%%%%%%%%%%%%%%%%%%%%%%%%%%%%%%%%%%%%%%%%%%%%%%%%%%%
The suggested \acp{EB} onset temperature domain is quite hot, but
where the \acp{LTE} recipes become invalid the temperatures required
to explain the actual \acp{EB} visibilities, especially in the \SiIV\
lines, must be substantially higher, even nearing the \acp{CE} limits.

The suggested domain is significantly hotter than the relative heating
obtained by \citetads{2009A&A...508.1469A} in their \acp{EB}
simulation.
However, more recent \acp{MHD} simulations of comparable reconnection
phenomena show larger heating (\citeads{2014ApJ...788L...2A};
\citeads{2015ApJ...798L..10L}), likely through magnetosonic wave
excitation and dissipation (\citeads{2007PhPl...14l2905L}). 
It is beyond the observational nature of this paper to discuss or
suggest how \acp{EB}s may magnetically heat dense photospheric gas to
the temperatures required for their visibilities, but clearly
\acp{EB}s represent an intriguing challenge for such endeavors.

The suggested domain is also very dense, but large densities may
indeed occur in reconnection events within the photosphere,
especially at shocks from the downward-directed jet in the observed
bi-modal pattern.

A final speculation concerns the line source functions.
All lines studied here are resonant-scattering in the ALC7
chromosphere (Fig.~\ref{fig:ALC7}), even \Halpha, and so are the
ultraviolet continua through bound-free scattering. 
However, in \acp{EB}s, especially in their later cooling phases, the
detour contribution in the third term in Eq.~\ref{eq:Sline} may
outweigh the second through recombination paths. 
A tell-tale is emission in the double \MgII\
triplet line between \MgII\ \hk, of which
examples are shown in \PaperIII.  
The high ionization energy of \MgIII\ (80\,eV, highest after Li and
Be) makes this likely by making \MgIII\ the population reservoir over
10\,000--40\,000\,K (Fig.~\ref{fig:CE-LTE}), enhancing
recombination from it.

%%%%%%%%%%%%%%%%%%%%%%%%%%%%%%%%%%%%%%%%%%%%%%%%%%%%%%%%%%%%%%%%%%%%%%%%%%%%
\section{Conclusion}  \label{sec:conclusion}
%%%%%%%%%%%%%%%%%%%%%%%%%%%%%%%%%%%%%%%%%%%%%%%%%%%%%%%%%%%%%%%%%%%%%%%%%%%%
The extinction estimates in Fig.~\ref{fig:extLTE} suggest that
\acp{EB}s have hydrogen densities of about 10$^{15}$\,cm$^{-3}$ and
temperatures in the 10\,000-20\,000\,K range, much hotter than all 
\acp{NLTE} modeling of \acp{EB}s so far but far cooler than
\acp{CE} modeling would suggest.

Of the various \acp{EB} diagnostics \Halpha\ is the most outstanding
thanks to its electron-sensitive broadening, huge elemental abundance,
and large excitation energy.
The latter two combine into exceedingly large high-temperature
extinction; the Stark broadening produces observable moustaches by
shifting the corresponding hot-gas emissivity from the fibril-obscured
core to the extended line wings. 
The large excitation energy also causes extinction and emissivity
memory for hotter moments through retarded recombination balancing.
It also causes \Halpha-wing transparency of cooler surrounding and
overlying gas, permitting unobstructed \acp{EB} viewing.

The \acp{AIA} 1600 and 1700\,\AA\ channels are also good \acp{EB}
diagnostics because at high temperature they are dominated by the
Balmer continuum which shares the \Halpha\ properties -- except Stark
broadening, but they do not need that because overlying fibrils (as
the pink one in Fig.~\ref{fig:HION}) are transparent at these
wavelengths. 
However, this imaging is defocused by bound-free scattering in
surrounding gas.

Clean \acp{EB} imaging as in \Halpha\ moustaches
is also furnished by the \SiIV\ lines in which the overlying fibril
canopy is also transparent while the Balmer-continuum irradiation by
\acp{EB} feet burns an opacity hole around them and bound-free 
surround scattering causes no defocus.

How to proceed?
The present state of \acp{EB} research resembles \CaII\ \KtwoV\ grain
research two decades ago.  
EBs are minuscule rare brightenings of the outer \Halpha\ wings that
are hard to explain, as evident from the large  but confused
literature (\citeads{2013JPhCS.440a2007R}). 
\KtwoV\ grains are minuscule (but ubiquitous) brightenings of the
inner \CaIIK\ wings that were then also hard to explain, as evident
from the large but confused literature at the time
(\citeads{1991SoPh..134...15R}). 
The \KtwoV\ puzzle was solved with the data-driven simulation of
\citetads{1997ApJ...481..500C}; 
the clincher was that the observed spectral grain evolution pattern is
so specific that its precise reproduction (on the cover of
\citeads{1994chdy.conf.....C}) 
implied undeniable simulation verification.

It seems obvious that the next step in \acp{EB} research must be
similar reproduction of the rich set of \acp{EB} visibilities in
diverse diagnostics through numerical \acp{MHD} simulation with
detailed spectrum synthesis. 
The observations in Papers I--IV provide sufficient constraints for
undeniable simulation verification.

I do not suggest \acp{LTE} as a viable approach for such
effort. 
I have used the assumption of Saha-Boltzmann lower-level populations
here to gauge the ballpark of \acp{EB} extinction values in the
initial hot and dense \acp{EB} onsets. 
These may be close to \acp{LTE} (Sect.~\ref{sec:temporal}) or yet
higher (Sect.~\ref{sec:SE}), but the evaluation of the aftermaths must
be \acp{non-E} to track the level populations during the further
evolution.

While \acp{non-E} 3D simulation including detailed \acp{non-E} 3D
radiative transfer remains too resource-demanding, a shortcut
simplification for single-snapshot \Halpha\ synthesis is to maintain
or slowly decay fractional hydrogen populations after their peak, or
simply set the $n\is2$ population with
Eq.~\ref{eq:peakfit}.

Observationally, \acp{EB}s present a perfect target for solar
telescopes to verify promised angular resolution. 
The clean viewing of \acp{EB} \Halpha\ moustaches
implies rendering of detail down to kilometer scales, far below the
photon path lengths that affect other optical diagnostics.
In addition, the intrinsic moustache contrast is high so that
resolving fine structure is feasible at moderate Strehl ratio.
At good seeing the spatial moustache resolution is therefore limited
only by the telescope aperture and quality. 

The 25\,km resolution at \Halpha\ and Strehl above 0.6
promised for the Daniel K. Inouye Solar Telescope
are commensurate with simulation resolutions. 
Thus, \acp{EB}s may become the first solar reconnection phenomenon
diagnosed and understood in full detail.

The time seems also ripe to exploit the full-time full-disk sampling
of \acp{EB} occurrences which \acp{AIA} delivers in its 1700\,\AA\
channel to use \acp{EB}s as pin-pointers of photospheric strong-field
reconnection events for charting active-region field reconfiguration,
This is likely yet more fruitful for \acp{FAF}s sampled likewise
by \acp{AIA}'s 1600\,\AA\ channel. 
They are \acp{EB} siblings with the added interest that they affect
the higher solar atmosphere and so provide direct
lower boundary constraints to coronal field evolution.

Finally, the onset-\acp{LTE} recipes given above may 
explain extraordinary \Halpha\ visibility in other sudden heating
events and their aftermaths.
They then similarly invalidate \acp{SE} modeling of such phenomena.

%%%%%%%%%%%%%%%%%%%%%%%%%%%%%%%%%%%%%%%%%%%%%%%%%%%%%%%%%%%%%%%%%%%%%%%%%%%%
\begin{acknowledgements}
I am much indebted to C.~Zwaan (1928--1999) who often suggested
\acp{EB} line formation as research topic.
The didactic nature of this paper bears his stamp H.~Uitenbroek
provided the newest version of RH.
Comments from J.~Leenaarts and the referee improved the presentation.
CHIANTI is a project of George Mason University, the University of
Michigan and the University of Cambridge.
As always, I made much use of \acp{NASA}'s \acp{ADS} and the SolarSoft
library.
\end{acknowledgements}

%%%%%%%%%%%%%%%%%%%%%%%%%%%%%%%%%%%%%%%%%%%%%%%%%%%%%%%%%%%%%%%%%%%%%%%%%%%%
%% references
\bibliographystyle{aa-note}
\bibliography{rjrfiles,adsfiles} 

\end{document}